\documentclass[twocolumn,american,english,aps,prx,intlimits,longbibliography]{revtex4-2}
\usepackage{lmodern}

\usepackage[T1]{fontenc}
\usepackage[latin9]{inputenc}
\usepackage{color}
\usepackage{babel}
\usepackage{amsmath}
\usepackage{amssymb}
\usepackage{graphicx}
\usepackage[unicode=true,pdfusetitle,
 bookmarks=true,bookmarksnumbered=false,bookmarksopen=false,
 breaklinks=false,pdfborder={0 0 0},pdfborderstyle={},backref=false,colorlinks=true]
 {hyperref}

\makeatletter
\usepackage{graphicx}
\usepackage{amsmath}
\hypersetup{colorlinks,allcolors=blue,breaklinks=true}

\makeatother

\begin{document}
\title{Extensive long-range entanglement at finite temperatures from a nonequilibrium
bias}
\author{Shachar Fraenkel and Moshe Goldstein}
\affiliation{Raymond and Beverly Sackler School of Physics and Astronomy, Tel Aviv
University, Tel Aviv 6997801, Israel}
\begin{abstract}
Thermal equilibrium states of local quantum many-body systems are
notorious for their spatially decaying correlations, which place severe
restrictions on the types of many-body entanglement structures that
may be observed at finite temperatures. These restrictions may however
be defied when an out-of-equilibrium steady state is considered instead.
In this paper, we study the entanglement properties of free fermions
on a one-dimensional lattice that contains a generic charge- and energy-conserving
noninteracting impurity, and that is connected at its edges to two
reservoirs with different equilibrium energy distributions. These
distributions may differ in either temperature, chemical potential,
or both, thereby inducing an external bias. We analytically derive
exact asymptotic expressions for several quantum information measures
-- the mutual information, its R\'enyi generalizations, and the
fermionic negativity -- that quantify the correlation and entanglement
between two subsystems located on opposite sides of the impurity.
We show that all these measures scale (to a leading order) linearly
with the overlap between one subsystem and the mirror image of the
other (upon reflection of the latter about the impurity), independently
of the distance between the subsystems. While a simple proportionality
relation between the negativity and R\'enyi versions of the mutual
information is observed to hold at zero temperature, it breaks down
at finite temperatures, suggesting that these quantities represent
strong long-range correlations of different origins. Our results generalize
previous findings that were limited to the case of a chemical-potential
bias at zero temperature, rigorously demonstrating that the effect
of long-range volume-law entanglement is robust at finite temperatures.
\end{abstract}
\maketitle

\section{Introduction}

Nonequilibrium condensed matter systems produce a rich landscape of
physical phenomena, its exploration ceaselessly revealing behaviors
that do not fall under the auspices of traditional equilibrium statistical
mechanics. As we strive to chart the contours of this landscape, we
are thus called to construct novel theoretical formalisms that could
effectively capture such peculiar effects. For quantum many-body systems,
a perspective that has emerged in recent years, rooted in quantum
information theory, has been immensely conducive to the formulation
of a theoretical language that could encompass far-from-equilibrium
physics. This perspective uses measures of the amount and the spread
of quantum information stored in the state of a system as probes for
nonequilibrium many-body phenomena, allowing to describe these phenomena
quantitatively, as well as to unify their description across different
theoretical models.

As a marked example for the success of this approach, the proneness
of a closed quantum system to reach local thermal equilibrium (or
to avoid it) after a nonequilibrium quench has been repeatedly shown
to be closely connected to the dynamics of entanglement and to the
entanglement structure of energy eigenstates \citep{PhysRevLett.111.127201,PhysRevLett.111.127205,doi:10.1146/annurev-conmatphys-031214-014726,doi:10.1080/00018732.2016.1198134,doi:10.1073/pnas.1703516114,Deutsch_2018,10.21468/SciPostPhys.4.3.017,RevModPhys.91.021001,Serbyn2021}.
Additionally, quantum information measures can be used to distinguish
integrable systems from nonintegrable ones. Following a quench, measures
of the correlation between two disjoint subsystems of an integrable
system will exhibit a transient peak even when the subsystems are
separated by a large distance, while in a nonintegrable system the
height of this peak practically vanishes already at a modest separation
\citep{Mesty_n_2017,10.21468/SciPostPhys.4.3.017,Alba_2019,PhysRevB.100.115150},
a clear signature of information scrambling \citep{Hayden_2007,Swingle2018,xu2023scrambling}.

These long-range quantum correlations within an integrable system
can be made stationary rather than transient, if one considers an
open system instead of a closed one. Several recent studies \citep{PhysRevX.9.021007,PhysRevLett.123.110601,PhysRevX.13.011045,fraenkel2022extensive,bernard2023exact}
have revealed examples of steady states of open free-fermion systems
in which the mutual information (MI, a correlation measure that will
be precisely defined below) between subsystems follows a volume law,
meaning that it scales linearly with the size of the subsystems in
question. Such scaling of the MI is exclusive to nonequilibrium systems:
for two subsystems comprising a system at thermal equilibrium, their
MI is known to adhere to an area law \citep{PhysRevLett.100.070502},
meaning that it scales linearly with the size of their shared boundary
(we note that at a zero-temperature ground state this area law may
be violated, but only to a logarithmic order in subsystem size \citep{PhysRevLett.96.010404,PhysRevLett.96.100503,Calabrese_disjoint_2009};
such a logarithmic violation has also been predicted for a nonequilibrium
state beyond zero temperature \citep{PhysRevA.89.032321,PhysRevB.107.075157}).

All of these recent studies considered the steady state of a system
that is subjected to an external chemical-potential bias, and that
is characterized by inhomogeneity: Refs.~\citep{PhysRevX.9.021007,PhysRevLett.123.110601}
considered fermions in a disordered potential, relating the scaling
of the MI to the phase transition of the 3D noninteracting Anderson
model; Refs.~\citep{PhysRevX.13.011045,bernard2023exact} examined
a quantum variant of the 1D symmetric simple exclusion process, where
fermions experience random hopping between lattice sites; and in Ref.~\citep{fraenkel2022extensive}
we looked at a 1D tight-binding model hosting a noninteracting impurity,
and investigated the zero-temperature correlations between subsystems
on opposite sides of the impurity. Refs.~\citep{PhysRevX.9.021007,fraenkel2022extensive}
have additionally shown that the volume-law scaling applies also to
the fermionic negativity, a genuine measure of quantum entanglement
between subsystems (a precise definition is given below).

In this context, Ref.~\citep{fraenkel2022extensive} is distinct
in that it considered the correlation between spatially separated
subsystems. We have shown there that the extensive scaling of the
MI and the negativity holds even in the long-range limit, where the
distance between the subsystems is much larger than their lengths,
provided that their distances from the impurity are similar. This
remarkable entanglement structure is generated by fermions occupying
the excess energy states of the edge reservoir with the higher chemical
potential, and it stems from the coherence between the reflected part
and the transmitted part of each fermion that is scattered off the
impurity.

The emergence of extensive long-range entanglement in the steady state
of an inhomogeneous free system is close in spirit to the aforementioned
transient MI peak in a post-quench homogeneous integrable system,
as both phenomena arise due to the absence of decoherence in the system,
and the consequent ability of particles (or quasi-particles) to carry
quantum information over large distances while unobstructed. It is
therefore natural to ponder to what extent the mechanism observed
in the case of Ref.~\citep{fraenkel2022extensive} is general, and
may lead to similar entanglement structures in steady states beyond
that particular case.

In this work, we make progress in this direction by studying the same
free-fermion model as in Ref.~\citep{fraenkel2022extensive}, but
considering a more general form of the external nonequilibrium bias.
Instead of assuming that the edge reservoirs are at zero temperature
with different chemical potentials, we relax the assumption by requiring
simply that the energy distributions in the two reservoirs are different.
This includes the possibility of finite-temperature reservoirs, and
in particular the steady state formed due to a temperature bias with
no net particle current.

We analytically derive the asymptotic forms of the MI and negativity
within the steady state, and find the same volume-law scaling as the
one observed in the zero-temperature case: the two quantities scale
linearly with the mirror-image overlap between the two subsystems;
that is, the number of overlapping lattice sites between one subsystem
and the mirror-image of the other, when reflected about the location
of the impurity. On our way there, we also obtain exact expressions
for families of quantities that generalize the MI and negativity,
which are collectively referred to as R\'enyi generalizations, and
which contain additional information on the correlation structure
of the steady state. Our results thus contribute also to the growing
body of analytical work on quantum correlations in many-body systems
containing impurities, both in and out of equilibrium \citep{Eisler_2012,PhysRevB.88.085413,10.21468/SciPostPhys.6.1.004,Gruber_2020,Mintchev2021,10.21468/SciPostPhys.12.1.011,PhysRevLett.128.090603,Capizzi_2023,horvath2023chargeresolved,CapizziFCS_2023,Gouraud_2023,10.21468/SciPostPhys.14.4.070,rylands2023transport}.

The paper is structured as follows. In Sec.~\ref{sec:Basic-definitions}
we present the definitions of the quantum information measures that
lie at the center of our analysis, and explain their importance in
general. In Sec.~\ref{sec:model} we describe the model of interest
and define the subsystems for which correlation measures are to be
computed. Sec.~\ref{sec:Results} contains our main analytical results,
as well as a discussion of their physical implications and an intuitive
explanation enabling their natural interpretation. The details of
the derivation of these results appear in Sec.~\ref{sec:Analytical-derivation},
which is supplemented by three technical appendices. We conclude the
paper with Sec.~\ref{sec:Summary-and-outlook}, which summarizes
our main messages and looks ahead at future directions of research.

\section{Review of relevant quantum information quantities\label{sec:Basic-definitions}}

The following section provides a brief review of the quantum information
quantities studied in this work, including their formal definitions
and their meanings in terms of many-body correlations. Throughout
this section, we will use the notation $\rho_{{\scriptscriptstyle X}}$
to refer to the reduced density matrix of some subsystem $X$, which
is the density matrix left following a partial trace of the system
density matrix over the degrees of freedom of $X^{c}$, the complement
of $X$.

We first define the von Neumann entropy of subsystem $X$ as
\begin{equation}
S_{X}=-{\rm Tr}\!\left[\rho_{{\scriptscriptstyle X}}\ln\rho_{{\scriptscriptstyle X}}\right],\label{eq:vNE-definition}
\end{equation}
and its $n$th R\'enyi entropy as
\begin{equation}
S_{X}^{\left(n\right)}=\frac{1}{1-n}\ln{\rm Tr}\!\left[\left(\rho_{{\scriptscriptstyle X}}\right)^{n}\right].\label{eq:Renyi-entropy-definition}
\end{equation}
These quantities satisfy $\lim_{n\to1}S_{X}^{\left(n\right)}=S_{X}$,
a relation which is commonly used to extract the von Neumann entropy,
as R\'enyi entropies tend to be more directly accessible, both in
terms of analytical computation (e.g., using a replica trick \citep{Calabrese_2004})
as well as in terms of experimental measurement (using $n$ physical
or virtual copies of the system, see e.g.~Refs\@.~\citep{PhysRevLett.109.020504,PhysRevLett.109.020505,Islam2015,PhysRevLett.120.050406,PhysRevA.99.062309}).
When the total state of the system is pure, the von Neumann and R\'enyi
entropies quantify the entanglement between $X$ and $X^{c}$ \citep{RevModPhys.81.865};
our current work, however, deals in general with states that are globally
mixed.

The correlation between any two disjoint subsystems $X_{1}$ and $X_{2}$
can be quantified through their mutual information (MI), defined as
\begin{equation}
{\cal I}=S_{X_{1}}+S_{X_{2}}-S_{X_{1}\cup X_{2}}.
\end{equation}
The MI is non-negative by its definition, and serves as a measure
of the total classical and quantum correlations between $X_{1}$ and
$X_{2}$ \citep{PhysRevA.72.032317}. In analogy to the von Neumann
entropy, the MI can be expressed as the $n\to1$ limit of the quantity
\begin{equation}
{\cal I}^{\left(n\right)}=S_{X_{1}}^{\left(n\right)}+S_{X_{2}}^{\left(n\right)}-S_{X_{1}\cup X_{2}}^{\left(n\right)},\label{eq:RMI-definition}
\end{equation}
to which we refer as the R\'enyi mutual information (RMI). The RMI
vanishes when $X_{1}$ and $X_{2}$ are completely uncorrelated (i.e.,
when $\rho_{{\scriptscriptstyle X_{1}\cup X_{2}}}=\rho_{{\scriptscriptstyle X_{1}}}\otimes\rho_{{\scriptscriptstyle X_{2}}}$),
but is generally problematic as a measure of correlations between
subsystems, as it lacks certain desirable properties: unlike the MI,
the RMI may increase under local operations (i.e., operations restricted
to either $X_{1}$ or $X_{2}$), and it can also be negative \citep{Kormos_2017,Scalet2021computablerenyi}.
However, the accessibility of the R\'enyi entropies makes the RMI
a useful tool in the extraction of the MI, and, as we will show, in
our case of interest the MI and the RMI follow a similar scaling law.

\begin{figure*}
\begin{centering}
\includegraphics[viewport=0bp 100bp 1280bp 670bp,clip,width=1\textwidth]{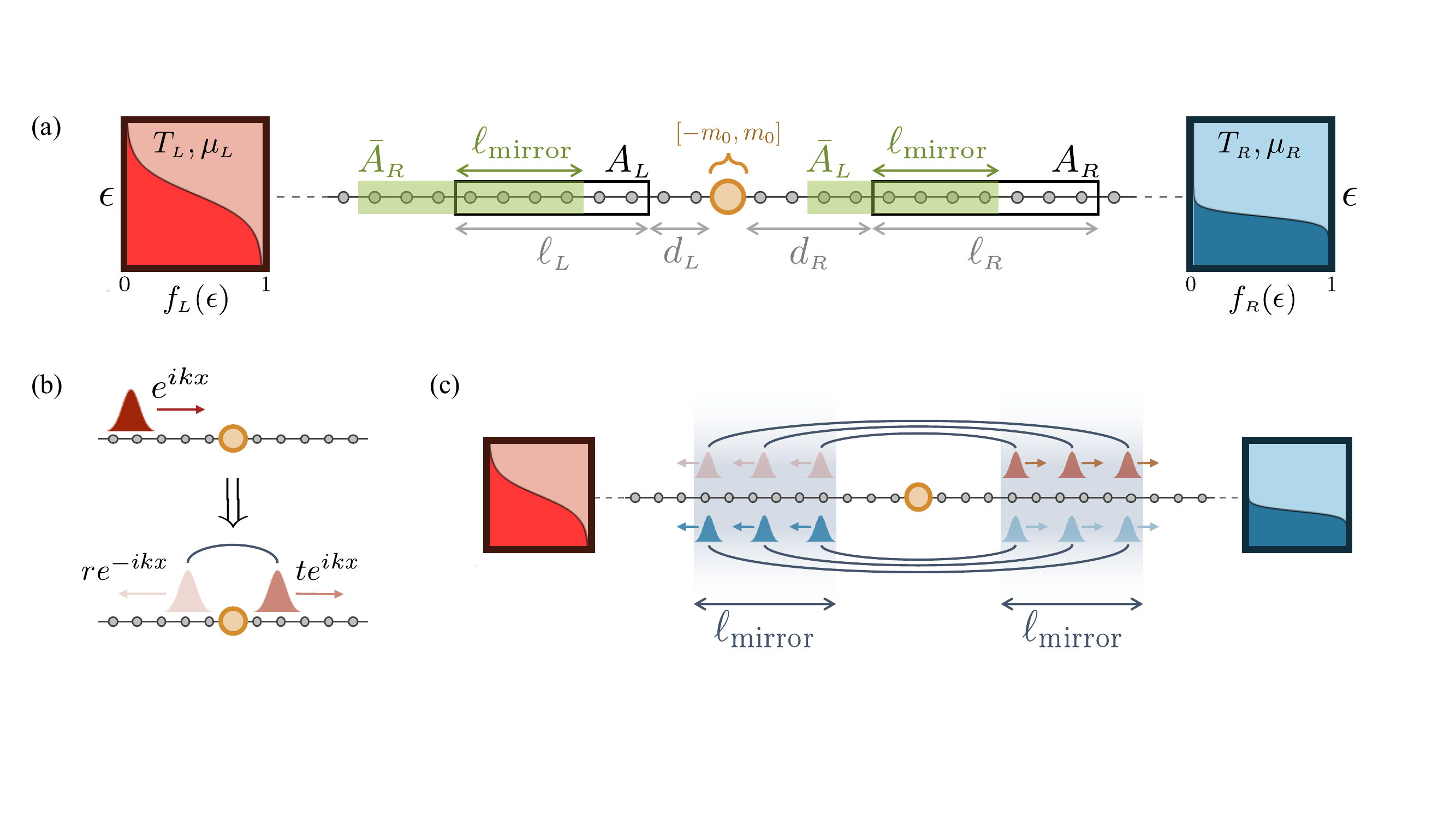}
\par\end{centering}
\caption{\label{fig:Model-sketch}(a) A schematic illustration of the model,
with an exemplary choice of subsystems. A long one-dimensional chain
of lattice sites is homogeneous almost everywhere, apart from a small
middle region (represented by the larger circle) which constitutes
a noninteracting impurity, and spans the sites with indices $-m_{0}\le m\le m_{0}$.
At its distant edges, the system is attached to two free-fermion reservoirs
with temperatures $T_{i}$ and chemical potentials $\mu_{i}$ ($i={\scriptstyle L,R}$),
giving rise to Fermi-Dirac energy distributions $f_{i}\!\left(\epsilon\right)$
(Eq.~(\ref{eq:Fermi-Dirac-distribution})). The subsystems $A_{{\scriptscriptstyle L}}$
and $A_{{\scriptscriptstyle R}}$, on opposite sides of the impurity,
are marked by black empty rectangles, with the length and distance
from the impurity of $A_{i}$ denoted by $\ell_{i}$ and $d_{i}$,
respectively. $\bar{A}_{i}=\left\{ m|-m\in A_{i}\right\} $ are the
mirror images of the subsystems with respect to the middle of the
chain, and $\ell_{{\rm mirror}}$ (defined in Eq.~(\ref{eq:ell-mirror-def}))
is the length of the overlap between one subsystem and the mirror
image of the other. (b) The homogeneity-breaking impurity induces
the scattering of an incoming wavepacket, concentrated around a certain
energy. This creates a transmitted part and a reflected part with
(generally energy-dependent) amplitudes $t$ and $r$ (respectively),
such that the two parts are coherently correlated. (c) At each energy
where, as a result of the nonequilibrium bias, there is an excess
of particles coming from one of the edge reservoirs, entanglement
between mirroring regions in the chain is generated due to the coherent
correlation between the counter-propagating scattered parts. The number
of wavepackets shared by the mirroring regions is proportional to
their length, leading to volume-law entanglement.}
\end{figure*}

There are also alternative definitions of index-dependent quantities
that generalize the MI, from which the latter can be obtained through
analytic continuation. One such quantity, which was recently studied
in the context of quantum field theories \citep{PhysRevLett.130.021603},
is the Petz-R\'enyi mutual information (PRMI):
\begin{equation}
{\cal D}^{\left(n\right)}=\frac{1}{n-1}\ln{\rm Tr}\!\left[\left(\rho_{{\scriptscriptstyle X_{1}\cup X_{2}}}\right)^{n}\left(\rho_{{\scriptscriptstyle X_{1}}}\otimes\rho_{{\scriptscriptstyle X_{2}}}\right)^{1-n}\right].\label{eq:PRMI-definition}
\end{equation}
The PRMI is a positive measure of correlations which satisfies $\lim_{n\to1}{\cal D}^{\left(n\right)}={\cal I}$.
It provides a set of bounds on the correlation function of any pair
of operators in which one operator is supported on $X_{1}$ and the
other is supported on $X_{2}$ \citep{PhysRevLett.130.021603,PhysRevLett.100.070502}.

Given that the MI does not distinguish between classical and quantum
correlations, it is desirable to study a measure that is sensitive
only to the latter. Such a measure is given by the fermionic negativity
\citep{PhysRevB.95.165101} (inspired by the logarithmic negativity
\citep{PhysRevA.65.032314,PhysRevLett.95.090503}), defined as
\begin{equation}
{\cal E}=\ln{\rm Tr}\sqrt{\left(\widetilde{\rho}_{{\scriptscriptstyle X_{1}\cup X_{2}}}\right)^{\dagger}\widetilde{\rho}_{{\scriptscriptstyle X_{1}\cup X_{2}}}},
\end{equation}
where the operation $\rho_{{\scriptscriptstyle X_{1}\cup X_{2}}}\rightarrow\widetilde{\rho}_{{\scriptscriptstyle X_{1}\cup X_{2}}}$
is the fermionic partial transpose (which is equivalent to a partial
time-reversal transformation, applied to either $X_{1}$ or $X_{2}$).
The negativity ${\cal E}$ measures the entanglement between $X_{1}$
and $X_{2}$ within any mixed state of $X_{1}\cup X_{2}$, and is
particularly useful when this state is Gaussian, as in this case its
computation is significantly simplified \citep{PhysRevB.95.165101,PhysRevB.97.165123,Shapourian_2019},
as we further explain in Sec.~\ref{sec:Analytical-derivation}. Again,
one can introduce an index-dependent quantity, the R\'enyi negativity,
which generically is easier to compute (as in our case, see Sec.~\ref{sec:Analytical-derivation})
and measure (see e.g.~Ref.~\citep{PhysRevA.99.062309}). The R\'enyi
negativity is defined for any even integer $n$ as
\begin{equation}
{\cal E}^{\left(n\right)}=\ln{\rm Tr}\!\left[\left(\left(\widetilde{\rho}_{{\scriptscriptstyle X_{1}\cup X_{2}}}\right)^{\dagger}\widetilde{\rho}_{{\scriptscriptstyle X_{1}\cup X_{2}}}\right)^{n/2}\right],
\end{equation}
and satisfies $\lim_{n\to1}{\cal E}^{\left(n\right)}={\cal E}$. Beyond
being auxiliary quantities for the computation of the negativity,
the R\'enyi negativities can be used to reconstruct the negativity
spectrum \citep{PhysRevB.94.195121,10.21468/SciPostPhys.7.3.037},
i.e., the spectrum of $\widetilde{\rho}_{{\scriptscriptstyle X_{1}\cup X_{2}}}$.
However, it is not itself a proper measure of correlations between
$X_{1}$ and $X_{2}$, as it may be nonzero even when these subsystems
are not correlated: if the state $\rho_{{\scriptscriptstyle X_{1}\cup X_{2}}}$
is separable, then ${\cal E}^{\left(n\right)}=\left(1-n\right)S_{X_{1}\cup X_{2}}^{\left(n\right)}$,
meaning that ${\cal E}^{\left(n\right)}$ is nonzero if $\rho_{{\scriptscriptstyle X_{1}\cup X_{2}}}$
is also mixed.

In the analysis that follows, we would like to plot our results for
the correlation measures relative to some meaningful scale. We therefore
mention here also the maximal possible values of the MI and negativity
between $X_{1}$ and $X_{2}$:
\begin{align}
{\cal I}_{{\rm max}} & =2\min\!\left\{ \left|X_{1}\right|,\left|X_{2}\right|\right\} \ln2,\nonumber \\
{\cal E}_{{\rm max}} & =\min\!\left\{ \left|X_{1}\right|,\left|X_{2}\right|\right\} \ln2,\label{eq:Maximal-MI-and-negativity}
\end{align}
with $\left|X_{i}\right|$ being the number of fermionic modes supported
in $X_{i}$ ($i=1,2$). The values in Eq\@.~(\ref{eq:Maximal-MI-and-negativity})
are the ones saturated by the state of maximal entanglement between
the subsystems, equivalent to them sharing $\min\!\left\{ \left|X_{1}\right|,\left|X_{2}\right|\right\} $
Bell pairs \citep{RevModPhys.81.865,PhysRevA.99.022310}. For any
$n$, the maximal possible values of ${\cal I}^{\left(n\right)}$
and ${\cal D}^{\left(n\right)}$ are also equal to ${\cal I}_{{\rm max}}$.

\section{The system and its steady state\label{sec:model}}

Here we define precisely the model and the nonequilibrium state that
are the subject of this work, and set up the notation with which we
will express our analytical results.

We consider a one-dimensional lattice, the sites of which can be occupied
by free spinless fermions. The lattice is connected at its edges to
two reservoirs of free fermions, both of which at equilibrium: the
left reservoir is characterized by a chemical potential $\mu_{{\scriptscriptstyle L}}$
and a temperature $T_{{\scriptscriptstyle L}}$, and the right reservoir
is characterized by a chemical potential $\mu_{{\scriptscriptstyle R}}$
and a temperature $T_{{\scriptscriptstyle R}}$. The lattice itself
is modeled as a tight-binding chain which is homogeneous almost everywhere.
The homogeneity of the chain is broken only in a small region near
its center, a region which constitutes a noninteracting impurity that
conserves the number of particles. The model is therefore governed
by a Hamiltonian of the form
\begin{equation}
{\cal H}=-\eta\!\sum_{m=m_{0}}^{\infty}\!\!\left[c_{m}^{\dagger}c_{m+1}+c_{-m}^{\dagger}c_{-m-1}+{\rm h.c.}\right]+{\cal H}_{{\rm scat}},\label{eq:Model-Hamiltonian}
\end{equation}
with $\eta>0$ being a real hopping amplitude, $c_{m}$ being the
fermionic annihilation operator associated with the site $m$, $m_{0}$
being some fixed integer, and ${\cal H}_{{\rm scat}}$ being the term
associated with the impurity. ${\cal H}_{{\rm scat}}$ is a combination
of quadratic terms of the form $c_{m}^{\dagger}c_{m'}$, with the
indices $m,m'$ being limited to the region $-m_{0}\le m,m'\le m_{0}$,
or possibly associated with additional sites attached to this region
from the side. Realistically, the lattice must be finite considering
its attachment to edge reservoirs, but the infinite sum appearing
in Eq.~(\ref{eq:Model-Hamiltonian}) is still a valid representation
if we interpret the correlation structure that we discover as pertaining
to a region around the impurity that is much smaller than the actual
length of the lattice.

The single-particle energy eigenbasis of the Hamiltonian in Eq.~(\ref{eq:Model-Hamiltonian})
is generically comprised of bound states, that are exponentially localized
at the impurity, and of extended states \citep{doi:10.1063/1.525968}.
The effect of the bound states is negligible when considering correlations
between subsystems that exclude the impurity, as we do in this work,
and we therefore ignore those states (this is justified even when
the distance between the subsystems and the impurity is small; we
are concerned with volume-law terms of correlation measures, while
the exponential decay of bound state wavefunctions means that they
can contribute at most to the subleading area-law terms). The form
of the extended states, on the other hand, is what gives rise to the
unusual entanglement structure that we find here. Outside the impurity
region, this form can be concisely represented by a unitary scattering
matrix $S\!\left(k\right)$ of size $2\times2$ \citep{merzbacher1998quantum},
defined for any $0<k<\pi$: 
\begin{equation}
S\!\left(k\right)=\left(\begin{array}{cc}
r_{{\scriptscriptstyle L}}\!\left(k\right) & t_{{\scriptscriptstyle R}}\!\left(k\right)\\
t_{{\scriptscriptstyle L}}\!\left(k\right) & r_{{\scriptscriptstyle R}}\!\left(k\right)
\end{array}\right).\label{eq:Scattering-matrix}
\end{equation}
The diagonal entries of $S\!\left(k\right)$ are reflection amplitudes,
and its off-diagonal entries are transmission amplitudes. We let ${\cal R}=\left|r_{{\scriptscriptstyle L}}\right|^{2}=\left|r_{{\scriptscriptstyle R}}\right|^{2}$
and ${\cal T}=\left|t_{{\scriptscriptstyle L}}\right|^{2}=\left|t_{{\scriptscriptstyle R}}\right|^{2}$
denote the reflection and transmission probabilities, respectively.
Since $S\!\left(k\right)$ is unitary, ${\cal T}\!\left(k\right)+{\cal R}\!\left(k\right)=1$.

The extended states $|k\rangle$ are parameterized by the momentum
coordinate $k$ with $0<\left|k\right|<\pi$, and are associated with
the single-particle energy spectrum $\epsilon\!\left(k\right)=-2\eta\cos\!k$.
They are comprised of left scattering states, defined for any $0<k<\pi$
by the wavefunction

\begin{equation}
\left\langle m|k\right\rangle =\begin{cases}
e^{ikm}+r_{{\scriptscriptstyle L}}\!\left(\left|k\right|\right)e^{-ikm} & m<-m_{0},\\
t_{{\scriptscriptstyle L}}\!\left(\left|k\right|\right)e^{ikm} & m>m_{0},
\end{cases}\label{eq:Left-scattering-states}
\end{equation}
together with right scattering states, given for any $-\pi<k<0$ by
\begin{equation}
\left\langle m|k\right\rangle =\begin{cases}
t_{{\scriptscriptstyle R}}\!\left(\left|k\right|\right)e^{ikm} & m<-m_{0},\\
e^{ikm}+r_{{\scriptscriptstyle R}}\!\left(\left|k\right|\right)e^{-ikm} & m>m_{0}.
\end{cases}\label{eq:Right-scattering-states}
\end{equation}
The form of the wavefunctions inside the impurity region is irrelevant
when considering correlations between subsystems that exclude this
region. Their form outside the impurity region clearly depends only
on the scattering matrix in Eq.~(\ref{eq:Scattering-matrix}) (we
emphasize that this form is exact only in the limit of an infinite
chain).

The many-body nonequilibrium steady state of the system is determined
by the single-particle energy distributions of the two edge reservoirs.
In each reservoir, a state with energy $\epsilon$ is occupied with
the probability $f_{i}\!\left(\epsilon\right)$ (with $i={\scriptstyle L,R}$),
which is given by the equilibrium Fermi-Dirac distribution:
\begin{equation}
f_{i}\!\left(\epsilon\right)=\frac{1}{\exp\left[\left(\epsilon-\mu_{i}\right)\!/T_{i}\right]+1}.\label{eq:Fermi-Dirac-distribution}
\end{equation}
In what follows, we will express these distributions as functions
of $k$, i.e., $f_{i}\!\left(k\right)=f_{i}\!\left(\epsilon\!\left(k\right)\right)$.
Within the lattice, an energy eigenstate $|k\rangle$ is therefore
occupied according to the distribution $\widetilde{f}\!\left(k\right)$,
defined as\foreignlanguage{american}{
\begin{equation}
\widetilde{f}\!\left(k\right)=\begin{cases}
f_{{\scriptscriptstyle R}}\!\left(k\right) & k<0,\\
f_{{\scriptscriptstyle L}}\!\left(k\right) & k>0.
\end{cases}
\end{equation}
In other words, if we let $c_{k}$ denote the annihilation operator
associated with the single-particle state $|k\rangle$, the nonequilibrium
steady state $\rho_{{\scriptscriptstyle {\rm NESS}}}$ of the system
is given by
\begin{equation}
\rho_{{\scriptscriptstyle {\rm NESS}}}=\prod_{k}\left[\widetilde{f}\!\left(k\right)c_{k}^{\dagger}c_{k}+\left(1-\widetilde{f}\!\left(k\right)\right)c_{k}c_{k}^{\dagger}\right],\label{eq:steady-state-definition}
\end{equation}
which is a Gaussian state.}

\selectlanguage{american}%
Throughout the rest of the paper, we address the correlations between
two subsystems, denoted by $A_{{\scriptscriptstyle L}}$ and $A_{{\scriptscriptstyle R}}$,
as measured by the different quantities reviewed in Sec.~\ref{sec:Basic-definitions}.
The interval $A_{{\scriptscriptstyle L}}$ is located to the left
of the impurity and spans the sites $m$ with $-d_{{\scriptscriptstyle L}}-\ell_{{\scriptscriptstyle L}}\le m+m_{0}\le-d_{{\scriptscriptstyle L}}-1$,
while the interval $A_{{\scriptscriptstyle R}}$ is located to the
right of the impurity and spans the sites $m$ with $d_{{\scriptscriptstyle R}}+1\le m-m_{0}\le d_{{\scriptscriptstyle R}}+\ell_{{\scriptscriptstyle R}}$.
Thus, $\ell_{i}$ denotes the number of sites in $A_{i}$, and $d_{i}$
represents the number of sites separating the impurity region from
the nearest boundary of $A_{i}$ \foreignlanguage{english}{($i={\scriptstyle L,R}$)}.
The length scales $\ell_{i}$ and $d_{i}$ are assumed to be much
larger than $2m_{0}+1,$ which is the size of the impurity.

As will become apparent when discussing our results, it is also useful
to introduce notations for the mirror images of these subsystems:
we let $\bar{A}_{i}=\left\{ m|-m\in A_{i}\right\} $ denote the set
of sites obtained by reflecting the sites in $A_{i}$ about the middle
site $m=0$, which also marks the center of the impurity. We can then
refer to either $\bar{A}_{{\scriptscriptstyle L}}\cap A_{{\scriptscriptstyle R}}$
or $A_{{\scriptscriptstyle L}}\cap\bar{A}_{{\scriptscriptstyle R}}$
as the mirror-image overlap between the two subsystems, i.e., the
overlap between one subsystem and the mirror image of the other. Finally,
we let $\ell_{{\rm mirror}}=\left|\bar{A}_{{\scriptscriptstyle L}}\cap A_{{\scriptscriptstyle R}}\right|=\left|A_{{\scriptscriptstyle L}}\cap\bar{A}_{{\scriptscriptstyle R}}\right|$
denote the length of this overlap, which is equivalent to defining
\begin{equation}
\ell_{{\rm mirror}}=\max\!\left\{ \min\!\left\{ d_{{\scriptscriptstyle L}}\!+\!\ell_{{\scriptscriptstyle L}},d_{{\scriptscriptstyle R}}\!+\!\ell_{{\scriptscriptstyle R}}\right\} \!-\!\max\!\left\{ d_{{\scriptscriptstyle L}},d_{{\scriptscriptstyle R}}\right\} ,0\right\} .\label{eq:ell-mirror-def}
\end{equation}
The length scale $\ell_{{\rm mirror}}$ is the one with respect to
which the correlation measures that we consider exhibit their volume-law
scaling. Importantly, it depends on the distances $d_{{\scriptscriptstyle L}}$
and $d_{{\scriptscriptstyle R}}$ of the subsystems from the impurity
only through $d_{{\scriptscriptstyle L}}-d_{{\scriptscriptstyle R}}$.

\selectlanguage{english}%
A schematic illustration of the system appears in Fig.~\ref{fig:Model-sketch}(a),
where we also show an example for a choice of $A_{{\scriptscriptstyle L}}$
and $A_{{\scriptscriptstyle R}}$, and indicate the associated length
scales. We reiterate the fact that the results that we present in
the following section are exact only in the limit of an infinite tight-binding
chain between the two reservoirs; mild corrections might appear when
the chain is long but finite. Alternatively, we may interpret these
results as applying to the infinite-time limit after a quench where
two semi-infinite chains with different initial equilibrium states
are joined together (or, in the case of long but finite chains, to
an intermediate time before the system edges are affected by the quench).

\section{Exact results\label{sec:Results}}

In this section we present the main results of our work, which are
the exact formulas for the leading-order asymptotics of the correlation
measures defined in Sec.~\ref{sec:Basic-definitions}, when applied
to the model and subsystems defined in Sec\@.~\ref{sec:model}.
We discuss the interpretation of these analytical formulas and their
various limits, and compare them to numerical results. The derivation
of the analytical results is detailed in Sec.~\ref{sec:Analytical-derivation}.

\begin{figure*}
\includegraphics[viewport=100bp 150bp 1250bp 800bp,clip,width=1\textwidth]{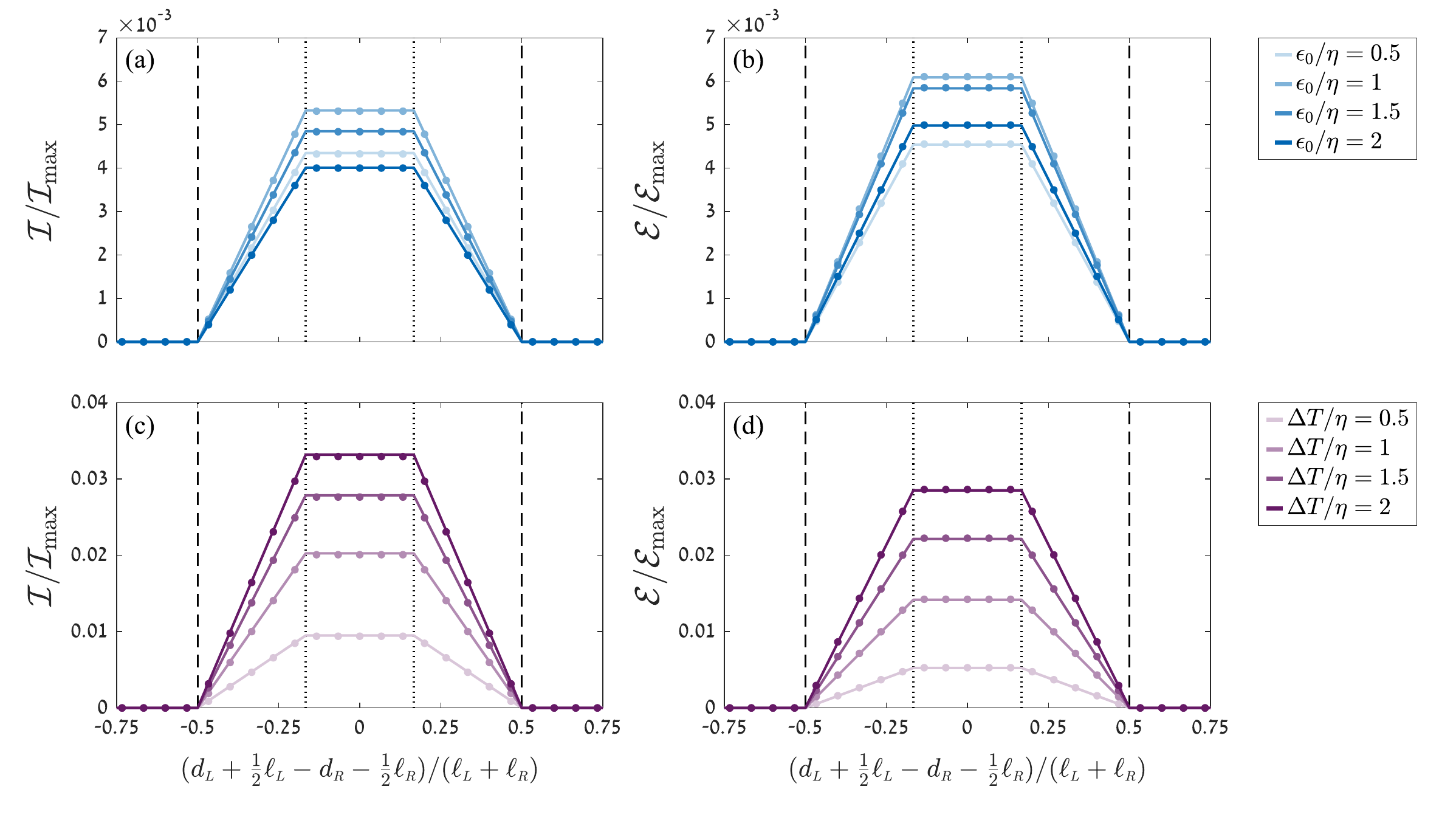}

\caption{\label{fig:Results-temperature-bias} The mutual information and negativity
between $A_{{\scriptscriptstyle L}}$ and $A_{{\scriptscriptstyle R}}$
in the resonant level model, under a temperature bias and with no
chemical-potential bias, setting $\mu_{{\scriptscriptstyle L}}=\mu_{{\scriptscriptstyle R}}=0$.
The MI and negativity are normalized by their maximal possible values
(see Eq.~(\ref{eq:Maximal-MI-and-negativity})). In all panels, the
subsystem lengths are fixed to be $\ell_{{\scriptscriptstyle L}}=100$
and $\ell_{{\scriptscriptstyle R}}=200$, while their relative distance
from the impurity $d_{{\scriptscriptstyle L}}-d_{{\scriptscriptstyle R}}$
is varied. Note that $d_{i}+\frac{1}{2}\ell_{i}$, appearing in the
label of the horizontal axis, is the distance from the midpoint of
$A_{i}$ to the impurity ($i={\scriptstyle L,R}$). Continuous lines
represent analytical results (computed from Eqs.~(\ref{eq:MI-main-result})
and (\ref{eq:Negativity-main-result}) for the MI and negativity,
respectively), while dots represent numerical results (computed in
the limit $d_{i}/\ell_{i}\to\infty$, as explained in Subsec.~\ref{subsec:Long-range-limit-correlations}).
In each panel, between the two vertical dotted lines the values of
$d_{{\scriptscriptstyle L}}-d_{{\scriptscriptstyle R}}$ are such
that $\bar{A}_{{\scriptscriptstyle L}}\subset A_{{\scriptscriptstyle R}}$
(i.e., maximal mirror-image overlap), and between the two vertical
dashed lines they are such that $\bar{A}_{{\scriptscriptstyle L}}\cap A_{{\scriptscriptstyle R}}\protect\neq\phi$
(i.e., nonzero mirror-image overlap). Top panels: (a) The MI and (b)
the negativity for different values of the impurity on-site energy
$\epsilon_{0}$, with $T_{{\scriptscriptstyle L}}=2\eta$ and $T_{{\scriptscriptstyle R}}=\eta$.
Bottom panels: (c) The MI and (d) the negativity for different values
of the temperature bias $\Delta T=T_{{\scriptscriptstyle L}}-T_{{\scriptscriptstyle R}}$,
with $T_{{\scriptscriptstyle R}}=0.5\eta$ and $\epsilon_{0}=\eta$.}
\end{figure*}

Our analysis shows that the $n$th RMI between $A_{{\scriptscriptstyle L}}$
and $A_{{\scriptscriptstyle R}}$ scales as
\begin{align}
{\cal I}^{\left(n\right)} & \sim\frac{\ell_{{\rm mirror}}}{1-n}\negthinspace\int_{0}^{\pi}\!\!\frac{dk}{2\pi}\bigg\{\!\!\ln\!\left[\left({\cal T}f_{{\scriptscriptstyle L}}\!+\!{\cal R}f_{{\scriptscriptstyle R}}\right)^{n}\!+\!\left(1\!-\!{\cal T}f_{{\scriptscriptstyle L}}\!-\!{\cal R}f_{{\scriptscriptstyle R}}\right)^{n}\right]\nonumber \\
 & +\ln\!\left[\left({\cal R}f_{{\scriptscriptstyle L}}\!+\!{\cal T}f_{{\scriptscriptstyle R}}\right)^{n}\!+\!\left(1\!-\!{\cal R}f_{{\scriptscriptstyle L}}\!-\!{\cal T}f_{{\scriptscriptstyle R}}\right)^{n}\right]\nonumber \\
 & -\ln\!\left[\left(f_{{\scriptscriptstyle L}}\right)^{n}+\left(1\!-\!f_{{\scriptscriptstyle L}}\right)^{n}\right]-\ln\!\left[\left(f_{{\scriptscriptstyle R}}\right)^{n}+\left(1\!-\!f_{{\scriptscriptstyle R}}\right)^{n}\right]\!\bigg\},\label{eq:Renyi-MI-main-result}
\end{align}
while the PRMI is given by\begin{widetext}
\begin{align}
{\cal D}^{\left(n\right)} & \sim\frac{\ell_{{\rm mirror}}}{n-1}\!\int_{0}^{\pi}\!\!\frac{dk}{2\pi}\ln\!\bigg[{\cal T}\!\left(\frac{\left(f_{{\scriptscriptstyle L}}\right)^{n}}{\left({\cal T}f_{{\scriptscriptstyle L}}\!+\!{\cal R}f_{{\scriptscriptstyle R}}\right)^{n-1}}+\frac{\left(1\!-\!f_{{\scriptscriptstyle L}}\right)^{n}}{\left(1\!-\!{\cal T}f_{{\scriptscriptstyle L}}\!-\!{\cal R}f_{{\scriptscriptstyle R}}\right)^{n-1}}\right)\!\left(\frac{\left(f_{{\scriptscriptstyle R}}\right)^{n}}{\left({\cal R}f_{{\scriptscriptstyle L}}\!+\!{\cal T}f_{{\scriptscriptstyle R}}\right)^{n-1}}+\frac{\left(1\!-\!f_{{\scriptscriptstyle R}}\right)^{n}}{\left(1\!-\!{\cal R}f_{{\scriptscriptstyle L}}\!-\!{\cal T}f_{{\scriptscriptstyle R}}\right)^{n-1}}\right)\nonumber \\
 & +{\cal R}\!\left(\frac{\left(f_{{\scriptscriptstyle L}}\right)^{n}}{\left({\cal R}f_{{\scriptscriptstyle L}}\!+\!{\cal T}f_{{\scriptscriptstyle R}}\right)^{n-1}}+\frac{\left(1\!-\!f_{{\scriptscriptstyle L}}\right)^{n}}{\left(1\!-\!{\cal R}f_{{\scriptscriptstyle L}}\!-\!{\cal T}f_{{\scriptscriptstyle R}}\right)^{n-1}}\right)\!\left(\frac{\left(f_{{\scriptscriptstyle R}}\right)^{n}}{\left({\cal T}f_{{\scriptscriptstyle L}}\!+\!{\cal R}f_{{\scriptscriptstyle R}}\right)^{n-1}}+\frac{\left(1\!-\!f_{{\scriptscriptstyle R}}\right)^{n}}{\left(1\!-\!{\cal T}f_{{\scriptscriptstyle L}}\!-\!{\cal R}f_{{\scriptscriptstyle R}}\right)^{n-1}}\right)\bigg].\label{eq:PRMI-main-result}
\end{align}
 By taking the limit $n\to1$ of either Eq.~(\ref{eq:Renyi-MI-main-result})
or Eq.~(\ref{eq:PRMI-main-result}), we obtain the MI, which reads
\begin{align}
{\cal I} & \sim\ell_{{\rm mirror}}\int_{0}^{\pi}\frac{dk}{2\pi}\bigg\{\!-\left({\cal T}f_{{\scriptscriptstyle L}}\!+\!{\cal R}f_{{\scriptscriptstyle R}}\right)\ln\!\left({\cal T}f_{{\scriptscriptstyle L}}\!+\!{\cal R}f_{{\scriptscriptstyle R}}\right)-\left(1\!-\!{\cal T}f_{{\scriptscriptstyle L}}\!-\!{\cal R}f_{{\scriptscriptstyle R}}\right)\ln\!\left(1\!-\!{\cal T}f_{{\scriptscriptstyle L}}\!-\!{\cal R}f_{{\scriptscriptstyle R}}\right)+f_{{\scriptscriptstyle L}}\ln f_{{\scriptscriptstyle L}}+\left(1\!-\!f_{{\scriptscriptstyle L}}\right)\ln\!\left(1\!-\!f_{{\scriptscriptstyle L}}\right)\nonumber \\
 & -\left({\cal R}f_{{\scriptscriptstyle L}}\!+\!{\cal T}f_{{\scriptscriptstyle R}}\right)\ln\!\left({\cal R}f_{{\scriptscriptstyle L}}\!+\!{\cal T}f_{{\scriptscriptstyle R}}\right)-\left(1\!-\!{\cal R}f_{{\scriptscriptstyle L}}\!-\!{\cal T}f_{{\scriptscriptstyle R}}\right)\ln\!\left(1\!-\!{\cal R}f_{{\scriptscriptstyle L}}\!-\!{\cal T}f_{{\scriptscriptstyle R}}\right)+f_{{\scriptscriptstyle R}}\ln f_{{\scriptscriptstyle R}}+\left(1\!-\!f_{{\scriptscriptstyle R}}\right)\ln\!\left(1\!-\!f_{{\scriptscriptstyle R}}\right)\!\bigg\}.\label{eq:MI-main-result}
\end{align}
\end{widetext}Furthermore, the fermionic negativity between the two
subsystems is found to scale as
\begin{align}
{\cal E} & \sim\ell_{{\rm mirror}}\int_{0}^{\pi}\frac{dk}{2\pi}\ln\!\bigg[f_{{\scriptscriptstyle L}}+f_{{\scriptscriptstyle R}}-2f_{{\scriptscriptstyle L}}f_{{\scriptscriptstyle R}}\nonumber \\
 & +\sqrt{\left[1-f_{{\scriptscriptstyle L}}-f_{{\scriptscriptstyle R}}+2f_{{\scriptscriptstyle L}}f_{{\scriptscriptstyle R}}\right]^{2}+4{\cal TR}\left(f_{{\scriptscriptstyle L}}-f_{{\scriptscriptstyle R}}\right)^{2}}\bigg].\label{eq:Negativity-main-result}
\end{align}
Thus, the measures in Eqs.~(\ref{eq:Renyi-MI-main-result})--(\ref{eq:Negativity-main-result})
all scale linearly with $\ell_{{\rm mirror}}$, signaling volume-law
long-range correlation and entanglement between the two subsystems
(recall that $\ell_{{\rm mirror}}$ depends only on $d_{{\scriptscriptstyle L}}-d_{{\scriptscriptstyle R}}$,
hence the use of the term ``long-range'').

\begin{figure*}
\includegraphics[viewport=100bp 150bp 1250bp 800bp,clip,width=1\textwidth]{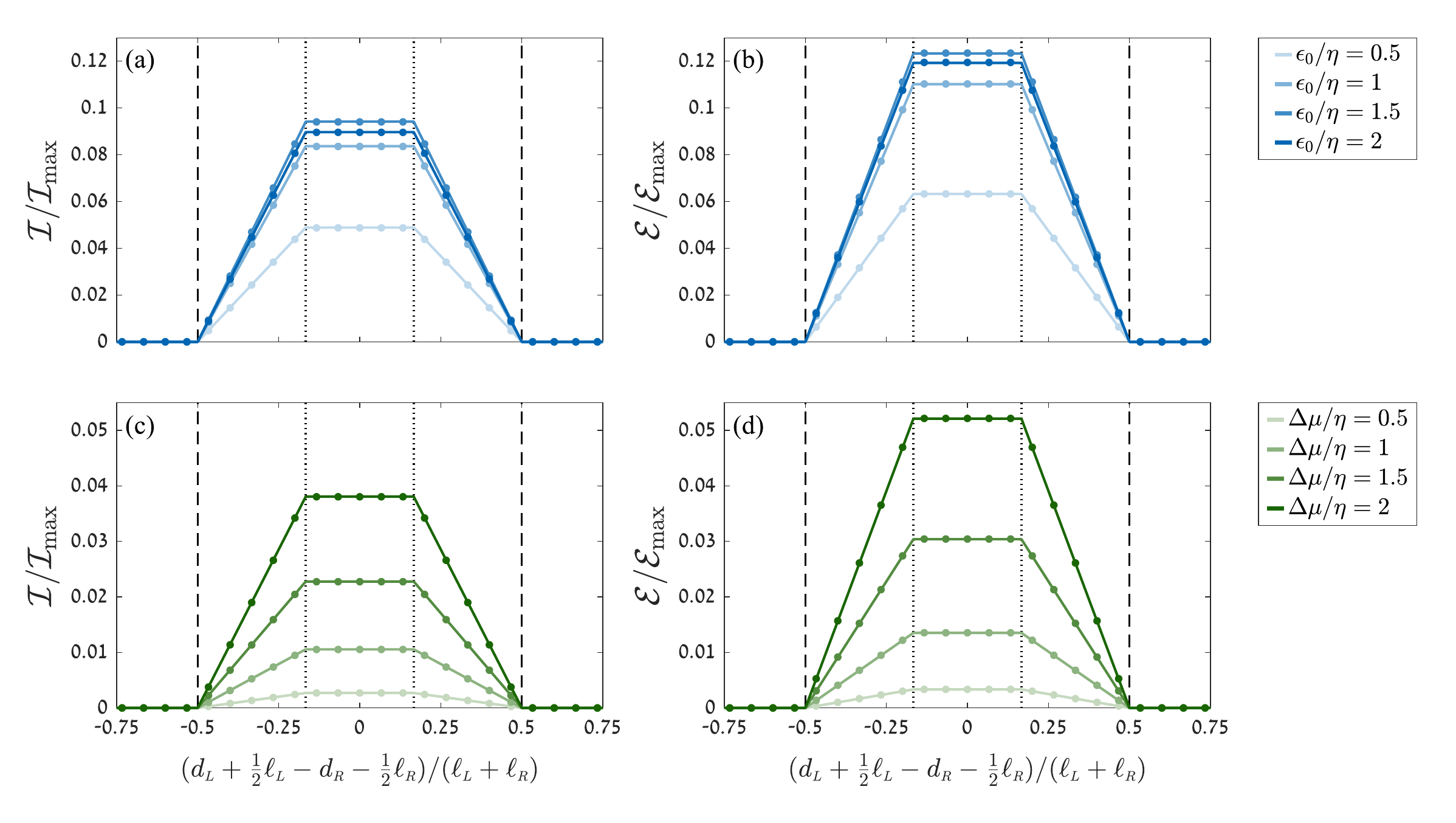}

\caption{\label{fig:Results-potential-bias}The mutual information and negativity
between $A_{{\scriptscriptstyle L}}$ and $A_{{\scriptscriptstyle R}}$
in the resonant level model, under a chemical-potential bias and with
no temperature bias, setting $T_{{\scriptscriptstyle L}}=T_{{\scriptscriptstyle R}}=\eta$.
The MI and negativity are normalized by their maximal possible values
(see Eq.~(\ref{eq:Maximal-MI-and-negativity})). In all panels, the
subsystem lengths are fixed to be $\ell_{{\scriptscriptstyle L}}=100$
and $\ell_{{\scriptscriptstyle R}}=200$, while their relative distance
from the impurity $d_{{\scriptscriptstyle L}}-d_{{\scriptscriptstyle R}}$
is varied. Note that $d_{i}+\frac{1}{2}\ell_{i}$, appearing in the
label of the horizontal axis, is the distance from the midpoint of
$A_{i}$ to the impurity ($i={\scriptstyle L,R}$). Continuous lines
represent analytical results (computed from Eqs.~(\ref{eq:MI-main-result})
and (\ref{eq:Negativity-main-result}) for the MI and negativity,
respectively), while dots represent numerical results (computed in
the limit $d_{i}/\ell_{i}\to\infty$, as explained in Subsec.~\ref{subsec:Long-range-limit-correlations}).
In each panel, between the two vertical dotted lines the values of
$d_{{\scriptscriptstyle L}}-d_{{\scriptscriptstyle R}}$ are such
that $\bar{A}_{{\scriptscriptstyle L}}\subset A_{{\scriptscriptstyle R}}$
(i.e., maximal mirror-image overlap), and between the two vertical
dashed lines they are such that $\bar{A}_{{\scriptscriptstyle L}}\cap A_{{\scriptscriptstyle R}}\protect\neq\phi$
(i.e., nonzero mirror-image overlap). Top panels: (a) The MI and (b)
the negativity for different values of the impurity on-site energy
$\epsilon_{0}$, with $\mu_{{\scriptscriptstyle L}}=-\mu_{{\scriptscriptstyle R}}=1.5\eta$.
Bottom panels: (c) The MI and (d) the negativity for different values
of the chemical-potential bias $\Delta\mu=\mu_{{\scriptscriptstyle L}}-\mu_{{\scriptscriptstyle R}}$,
with $\mu_{{\scriptscriptstyle R}}=0$ and $\epsilon_{0}=\eta$.}
\end{figure*}

We may immediately observe another property that these measures all
share: their volume-law terms vanish if, for all momenta $k\in\left[0,\pi\right]$,
either $f_{{\scriptscriptstyle L}}\!\left(k\right)=f_{{\scriptscriptstyle R}}\!\left(k\right)$
or ${\cal T}\!\left(k\right)\in\left\{ 0,1\right\} $. The first condition
corresponds to the absence of a nonequilibrium bias at the energy
$\epsilon\!\left(k\right)$, while the second translates into the
impurity being trivial, meaning perfectly transmissive or perfectly
reflective, at that energy. In contrast, if these conditions fail
to apply (for a nonzero-measure subset of $\left[0,k\right]$) --
i.e., in the presence of both a nonequilibrium bias and nontrivial
scattering -- then the volume-law terms in Eqs.~(\ref{eq:Renyi-MI-main-result})--(\ref{eq:Negativity-main-result})
are all positive. The bias and the nontrivial scattering by the impurity
are therefore necessary and sufficient conditions for the emergence
of the observed strong long-range entanglement in the steady state.

These two conditions together also render the functions ${\cal T}f_{{\scriptscriptstyle L}}\!+\!{\cal R}f_{{\scriptscriptstyle R}}$
and ${\cal R}f_{{\scriptscriptstyle L}}\!+\!{\cal T}f_{{\scriptscriptstyle R}}$
incompatible with equilibrium distributions; that is, none of them
can be written using an effective temperature and an effective chemical
potential. These two functions are featured in Eqs.~(\ref{eq:Renyi-MI-main-result})--(\ref{eq:MI-main-result})
as the arguments of entropic functions (e.g., $-x\ln x-\left(1-x\right)\ln\!\left(1-x\right)$
in Eq.~(\ref{eq:MI-main-result})) that otherwise tend to arise when
calculating entropic measures for some equilibrium state, with the
appropriate equilibrium distribution as their argument. The nonequilibrium
nature of the steady-state correlations is thus made conspicuous by
the mathematical structure of Eqs.~(\ref{eq:Renyi-MI-main-result})--(\ref{eq:MI-main-result}).

\begin{figure*}
\includegraphics[viewport=120bp 450bp 1250bp 800bp,clip,width=1\textwidth]{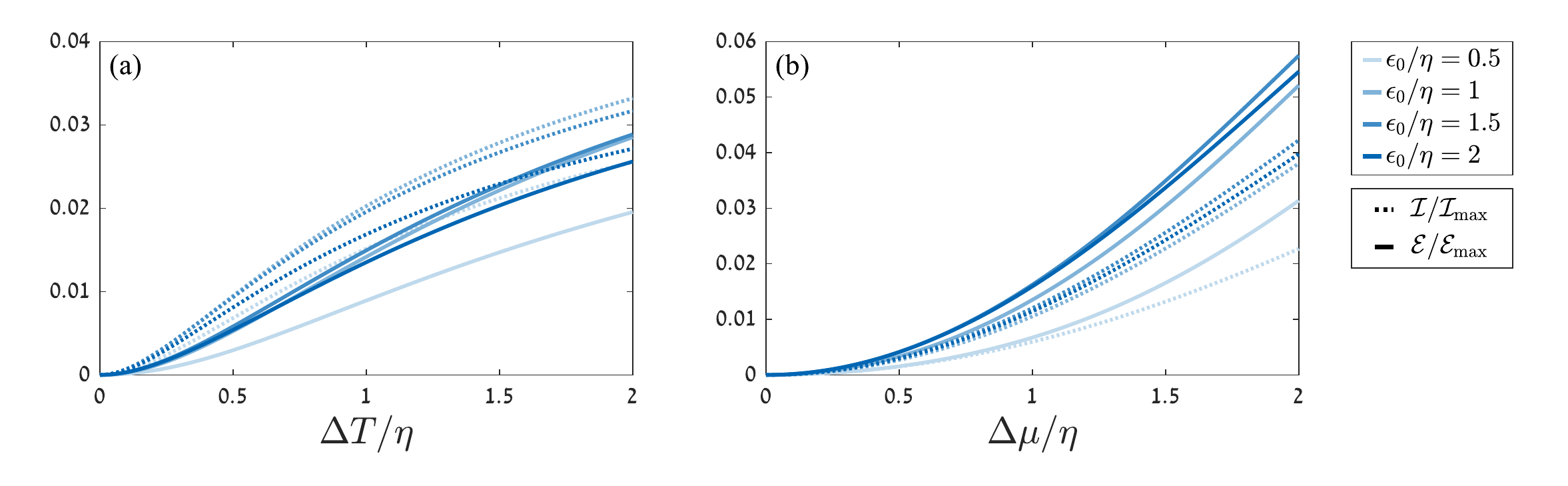}

\caption{\label{fig:Results-MI-Negativity-comparison}The mutual information
and negativity between $A_{{\scriptscriptstyle L}}$ and $A_{{\scriptscriptstyle R}}$
in the resonant level model, for a symmetric subsystem configuration
with $\ell_{{\scriptscriptstyle L}}=\ell_{{\scriptscriptstyle R}}=\ell_{{\rm mirror}}$.
The MI (dotted lines) and negativity (continuous lines) are computed
analytically (using Eqs.~(\ref{eq:MI-main-result})--(\ref{eq:Negativity-main-result}))
and normalized by their maximal possible values (see Eq.~(\ref{eq:Maximal-MI-and-negativity})),
such that the plotted values are independent of $\ell_{{\rm mirror}}$.
They are plotted for different values of the impurity on-site energy
$\epsilon_{0}$, as a function of (a) the temperature bias $\Delta T=T_{{\scriptscriptstyle L}}-T_{{\scriptscriptstyle R}}$
(having fixed $T_{{\scriptscriptstyle R}}=0.5\eta$ and $\mu_{{\scriptscriptstyle L}}=\mu_{{\scriptscriptstyle R}}=0$),
and of (b) the chemical-potential bias $\Delta\mu=\mu_{{\scriptscriptstyle L}}-\mu_{{\scriptscriptstyle R}}$
(having fixed $\mu_{{\scriptscriptstyle R}}=0$ and $T_{{\scriptscriptstyle L}}=T_{{\scriptscriptstyle R}}=\eta$).}
\end{figure*}

As in the zero-temperature case, the origin of the volume-law long-range
correlations can be interpreted by considering that, at each energy
participating in the nonequilibrium bias, there is an excess of particles
incoming from one reservoir rather than the other. A wavepacket representing
each such particle splits at the impurity into a coherent superposition
of two counter-propagating wavepackets, with the superposition amplitudes
determined by the scattering matrix (illustrated in Fig.~\ref{fig:Model-sketch}(b)).
With the two wavepackets moving uninterruptedly with group velocities
of the same magnitude, they generate entanglement between faraway
sites that have an equal distance from the impurity. This effect would
have been canceled within the many-body state if the same current
of particles at that energy had flowed in the opposite direction,
yet the nonequilibrium bias ensures that this is not so. Given the
stationary rate of incoming particles, mirroring regions share such
correlated wavepackets in a manner proportional to their length (illustrated
in Fig.~\ref{fig:Model-sketch}(c)). The only difference between
this interpretation and the more specific picture of the zero-temperature
case is simply that, in the latter, there is a certain window of energies
in which particles emerge only from one of the reservoirs, while in
the former particles possibly emerge from both reservoirs and still
generate entanglement due to an occupation bias.

There is, however, a noteworthy quantitative difference in the results
between the zero-temperature case and the more general scenario. By
plugging $T_{{\scriptscriptstyle L}}=T_{{\scriptscriptstyle R}}=0$
into Eqs.~(\ref{eq:Renyi-MI-main-result}) and (\ref{eq:Negativity-main-result}),
we recover the results that were reported in Ref.~\citep{fraenkel2022extensive}
for the RMI and the negativity. That is, letting $k_{+}=\max\!\left\{ k_{{\scriptscriptstyle F,L}},k_{{\scriptscriptstyle F,R}}\right\} $
and $k_{-}=\min\!\left\{ k_{{\scriptscriptstyle F,L}},k_{{\scriptscriptstyle F,R}}\right\} $
denote the Fermi momenta of the two reservoirs (with $k_{{\scriptscriptstyle F,}i}=\epsilon^{-1}\!\left(\mu_{i}\right)$),
we obtain 
\begin{equation}
{\cal I}^{\left(n\right)}\sim\frac{\ell_{{\rm mirror}}}{1-n}\negthinspace\int_{k_{-}}^{k_{+}}\!\frac{dk}{\pi}\!\ln\!\left[{\cal T}^{n}+{\cal R}^{n}\right],
\end{equation}
and find that the negativity and the $n=1/2$ RMI satisfy (to the
leading order) the simple relation ${\cal E}\sim\frac{1}{2}{\cal I}^{\left(1/2\right)}$.
This relation is observed in (interacting or noninteracting) integrable
systems following zero-temperature quenches, which are described by
the quasi-particle picture \citep{Alba_2019}, and in the post-quench
early-time dynamics of local systems in general \citep{PhysRevLett.129.140503}.
The relation ${\cal E}\sim\frac{1}{2}{\cal I}^{\left(1/2\right)}$
would have stemmed trivially from the definitions of the two quantities
had $A_{{\scriptscriptstyle L}}\cup A_{{\scriptscriptstyle R}}$ been
in a pure state, but is otherwise nontrivial \citep{PhysRevA.99.022310}.
It may be interpreted as indicating that the correlations between
the two subsystems are, to the leading order, purely quantum \citep{PhysRevLett.129.140503}.

In contrast, as can be seen from a direct examination of Eqs.~(\ref{eq:Renyi-MI-main-result})
and (\ref{eq:Negativity-main-result}), this relation between the
negativity and the $\frac{1}{2}$-RMI breaks down if at least one
of the reservoirs is at a finite temperature. A similar breakdown
of this relation was observed in nonequilibrium scenarios involving
dissipation \citep{alba2022logarithmic,Caceffo_2023}, as well as
in a steady state ensuing from a finite-temperature quench \citep{PhysRevB.107.075157}.
In a similar vein, at zero temperature we find that the result for
the PRMI in Eq.~(\ref{eq:PRMI-main-result}) is simply reduced to
${\cal D}^{\left(n\right)}\sim{\cal I}^{\left(3-2n\right)}$, also
implying that ${\cal E}\sim\frac{1}{2}{\cal D}^{\left(5/4\right)}$;
again, such relations would have been satisfied trivially had the
state of $A_{{\scriptscriptstyle L}}\cup A_{{\scriptscriptstyle R}}$
been pure \citep{PhysRevLett.130.021603}. This simple proportionality
relation of the PRMI to either the RMI or the negativity again breaks
down at finite temperatures.

\begin{figure*}
\includegraphics[viewport=100bp 150bp 1250bp 800bp,clip,width=1\textwidth]{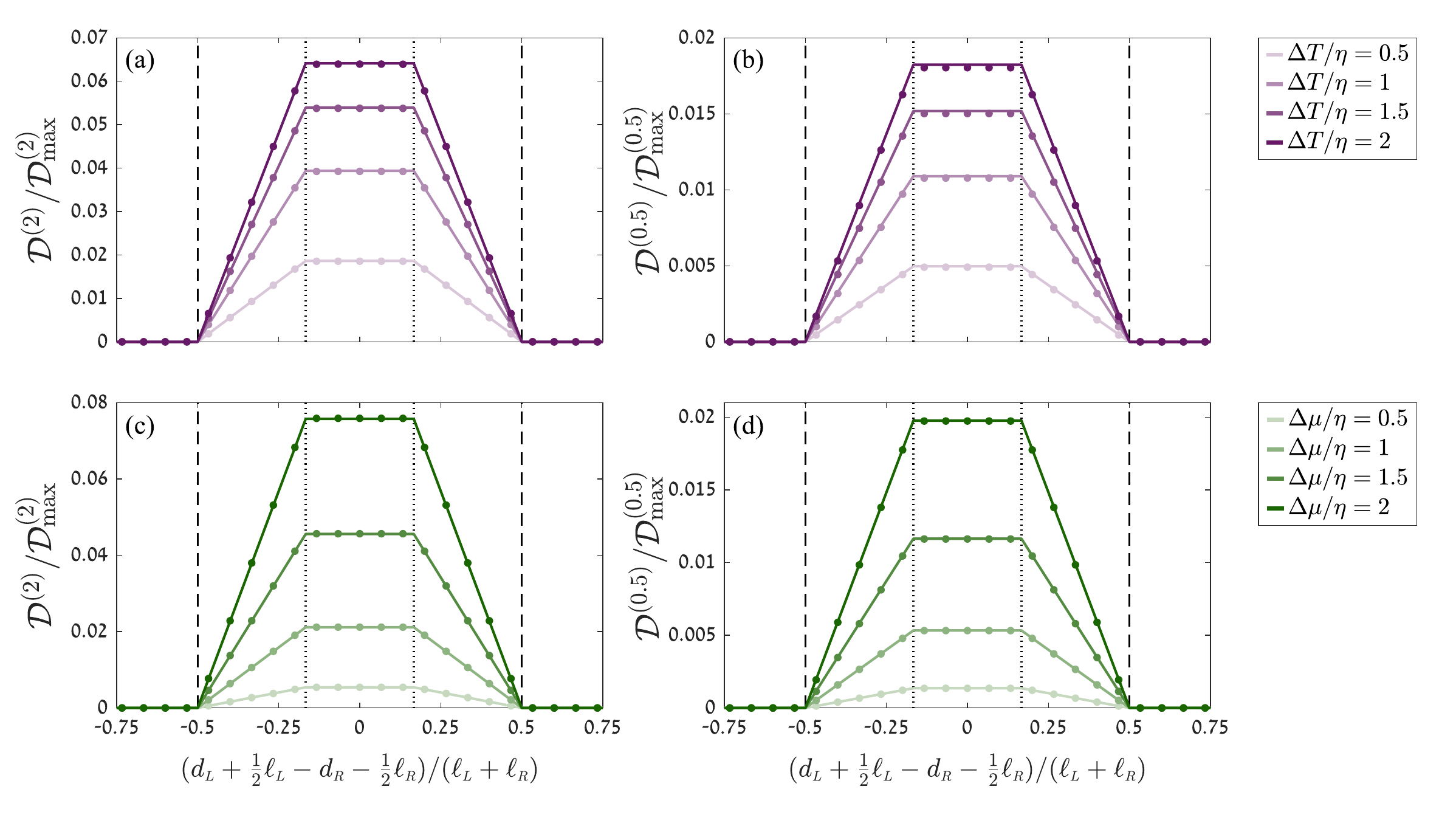}

\caption{\label{fig:Results-PRMI}The Petz-R\'enyi mutual information between
$A_{{\scriptscriptstyle L}}$ and $A_{{\scriptscriptstyle R}}$ in
the resonant level model, setting $\epsilon_{0}=\eta$. The PRMI is
normalized by its maximal possible value ${\cal D}_{{\rm max}}^{\left(n\right)}$,
equal to that of the MI (see Eq.~(\ref{eq:Maximal-MI-and-negativity})).
In all panels, the subsystem lengths are fixed to be $\ell_{{\scriptscriptstyle L}}=100$
and $\ell_{{\scriptscriptstyle R}}=200$, while their relative distance
from the impurity $d_{{\scriptscriptstyle L}}-d_{{\scriptscriptstyle R}}$
is varied. Note that $d_{i}+\frac{1}{2}\ell_{i}$, appearing in the
label of the horizontal axis, is the distance from the midpoint of
$A_{i}$ to the impurity ($i={\scriptstyle L,R}$). Continuous lines
represent analytical results (computed from Eq.~(\ref{eq:PRMI-main-result})),
while dots represent numerical results (computed in the limit $d_{i}/\ell_{i}\to\infty$,
as explained in Subsec.~\ref{subsec:Long-range-limit-correlations}).
In each panel, between the two vertical dotted lines the values of
$d_{{\scriptscriptstyle L}}-d_{{\scriptscriptstyle R}}$ are such
that $\bar{A}_{{\scriptscriptstyle L}}\subset A_{{\scriptscriptstyle R}}$
(i.e., maximal mirror-image overlap), and between the two vertical
dashed lines they are such that $\bar{A}_{{\scriptscriptstyle L}}\cap A_{{\scriptscriptstyle R}}\protect\neq\phi$
(i.e., nonzero mirror-image overlap). Top panels: The PRMI ${\cal D}^{\left(n\right)}$
for (a) $n=2$ and (b) $n=0.5$ in the case of a temperature bias
and with no chemical-potential bias. Results are plotted for different
values of the temperature bias $\Delta T=T_{{\scriptscriptstyle L}}-T_{{\scriptscriptstyle R}}$,
with $T_{{\scriptscriptstyle R}}=0.5\eta$ and $\mu_{{\scriptscriptstyle L}}=\mu_{{\scriptscriptstyle R}}=0$.
Bottom panels: The PRMI ${\cal D}^{\left(n\right)}$ for (c) $n=2$
and (d) $n=0.5$ in the case of a chemical-potential bias and with
no temperature bias. Results are plotted for different values of the
chemical-potential bias $\Delta\mu=\mu_{{\scriptscriptstyle L}}-\mu_{{\scriptscriptstyle R}}$,
with $\mu_{{\scriptscriptstyle R}}=0$ and $T_{{\scriptscriptstyle L}}=T_{{\scriptscriptstyle R}}=\eta$.}
\end{figure*}

To illustrate our results, in Figs.~\ref{fig:Results-temperature-bias}--\ref{fig:Results-potential-bias}
we plot the analytical expressions for the MI and negativity (Eqs.~(\ref{eq:MI-main-result})
and (\ref{eq:Negativity-main-result}), respectively) for a specific
choice of an impurity model. We focus on a resonant level model, substituting
$m_{0}=0$ and ${\cal H}_{{\rm scat}}=\epsilon_{0}c_{0}^{\dagger}c_{0}$
into Eq.~(\ref{eq:Model-Hamiltonian}), such that only the middle
site of the chain constitutes the impurity. It is straightforward
to check that in this case the momentum-dependent transmission probability
is given by 
\begin{equation}
{\cal T}\!\left(k\right)=\frac{\sin^{2}\!k}{\sin^{2}\!k+\left(\epsilon_{0}/2\eta\right)^{2}}.
\end{equation}
As a corroboration of our analytical calculation, we also plot numerical
results for the MI and negativity. These are computed in the long-range
limit $d_{i}/\ell_{i}\to\infty$ with $d_{{\scriptscriptstyle L}}-d_{{\scriptscriptstyle R}}$
kept fixed (a limit by which the analytical expressions are unaffected),
via an exact diagonalization of the two-point correlation matrix restricted
to $A_{{\scriptscriptstyle L}}\cup A_{{\scriptscriptstyle R}}$, as
we further elaborate in Subsec.~\ref{subsec:Long-range-limit-correlations}.

Fig.~\ref{fig:Results-temperature-bias} shows the results for a
pure temperature bias, where we plot the MI and negativity for a fixed
bias and different values of the impurity on-site energy $\epsilon_{0}$,
and vice versa. In Fig.~\ref{fig:Results-potential-bias} we present
a similar analysis, only with a pure chemical-potential bias at a
finite temperature instead of a temperature bias. In both figures,
we fix the subsystem lengths $\ell_{{\scriptscriptstyle L}}$ and
$\ell_{{\scriptscriptstyle R}}$ and vary $d_{{\scriptscriptstyle L}}-d_{{\scriptscriptstyle R}}$,
going between no mirror-image overlap and maximal mirror-image overlap.
We find an excellent agreement between the numerical results and the
analytical leading-order asymptotics.

In both figures, one may easily observe that a larger nonequilibrium
bias typically leads to stronger correlations, which is indeed natural
considering our interpretation of the results. The dependence on the
impurity on-site energy $\epsilon_{0}$ is more intricate, since a
change in $\epsilon_{0}$ affects the transmission probability across
the entire energy spectrum. Interestingly, by examining the top panels
of Fig.~\ref{fig:Results-temperature-bias}, one sees that it is
possible to find two values of $\epsilon_{0}$ such that the MI obtained
for one is larger than the MI obtained for the other, while the negativity
in the former case is smaller than in the latter. Such ``non-monotonicity''
between the MI and the negativity is a finite temperature effect,
given that at zero temperature it is forbidden due to the simple relation
between the negativity and the $\frac{1}{2}$-RMI. It is therefore
a manifestation of the MI capturing strong long-range correlations
beyond the quantum coherent correlations captured by the negativity.

It is clear that in our system both the quantum and classical correlations
have volume-law long-range behavior. However, the precise quantitative
separation of quantum correlations from classical ones tends to be
an arduous task, requiring the use of optimization-based measures
\citep{Henderson_2001,PhysRevLett.88.017901,RevModPhys.84.1655,Huang_2014},
and thus lies beyond the scope of this work. A certain flavor of this
issue may be noticed by superimposing the analytical results for the
MI and the negativity, each divided by its respective value for a
maximally entangled state; this is what we do in Fig.~\ref{fig:Results-MI-Negativity-comparison},
where both measures are plotted as functions of the temperature bias
or the chemical-potential bias, for a symmetric subsystem configuration
($\ell_{{\scriptscriptstyle L}}=\ell_{{\scriptscriptstyle R}}=\ell_{{\rm mirror}}$).
While the (normalized) MI is larger than the (normalized) negativity
in the case of a temperature bias, the opposite is true in the case
of the chemical-potential bias, suggesting that quantum correlations
are relatively stronger in the latter case. Yet, as we are not aware
of a definitive way for weighing the two types of correlations against
each other using the measures that we computed, this last observation
should not be construed as a quantitative statement.

Finally, we note that the derivation of the analytical formula for
the PRMI, given in Eq.~(\ref{eq:PRMI-main-result}), relies on a
recent conjecture \citep{10.21468/SciPostPhys.15.3.089} relating
to the asymptotics of a certain class of determinants (see details
in Subsec.~\ref{subsec:Asymptotics-of-PRMI-derivation}). This fact
calls for a verification of Eq.~(\ref{eq:PRMI-main-result}), by
comparing it to an independent numerical computation. In Fig.~\ref{fig:Results-PRMI}
we plot ${\cal D}^{\left(n\right)}$ for two values of $n$ and for
various instances of a nonequilibrium bias, finding in all cases that
Eq.~(\ref{eq:PRMI-main-result}) agrees nicely with the numerical
result.

\section{Analytical derivation\label{sec:Analytical-derivation}}

The following section describes the derivation of the results appearing
in Sec.~\ref{sec:Results}. In Subsec.~\ref{subsec:Two-point-correlations}
we recall general useful expressions for quantum information measures
in terms of two-point correlation functions, that apply when the many-body
state is Gaussian. Regarding the correlation function between sites
in $A_{{\scriptscriptstyle L}}\cup A_{{\scriptscriptstyle R}}$ for
our state of interest, we also recall its limit when the two subsystems
are distant (originally derived in Ref.~\citep{fraenkel2022extensive}),
and explain the importance of this limit within our calculations.
Subsequently, in Subsec.~\ref{subsec:Asymptotics-of-entropies-derivation}
we derive the forms of the R\'enyi entropies composing (via Eq.~(\ref{eq:RMI-definition}))
the RMI, leading to Eq.~(\ref{eq:Renyi-MI-main-result}); in Subsec.~\ref{subsec:Asymptotics-of-negativities-derivation}
we derive the form of the negativity, given in Eq.~(\ref{eq:Negativity-main-result});
and in Subsec.~\ref{subsec:Asymptotics-of-PRMI-derivation} we derive
the form of the PRMI, reported in Eq.~(\ref{eq:PRMI-main-result}).

\subsection{Reduction to two-point correlations\label{subsec:Two-point-correlations}}

The analysis we present in this paper relies on the fact that, for
any fermionic Gaussian many-body state (such as the one studied here),
subsystem entropies and negativities are fully determined by the spectrum
of two-point correlation matrices restricted to the subsystems of
interest. This amounts to an exponential reduction of the numerical
computational cost, and in many cases also allows to derive analytical
expressions through the use of asymptotic techniques related to the
structure of these correlation matrices.

For the steady state defined in Eq.~(\ref{eq:steady-state-definition}),
the two-point correlation function between any two sites can be written
as\foreignlanguage{american}{
\begin{equation}
\left\langle c_{j}^{\dagger}c_{m}\right\rangle =\int_{-\pi}^{\pi}\frac{dk}{2\pi}\widetilde{f}\!\left(k\right)\left\langle k|j\right\rangle \left\langle m|k\right\rangle .\label{eq:Correlation-function}
\end{equation}
A similar expression for the two-point correlation matrix applies
to any Gaussian state, upon determining the appropriate single-particle
states $|k\rangle$ that diagonalize the density matrix and their
corresponding occupation factors $\widetilde{f}\!\left(k\right)$.}

\subsubsection{General relations for fermionic Gaussian states}

\selectlanguage{american}%
Let $C_{X}$ denote the two-point correlation matrix restricted to
a subsystem $X$, meaning that its entries are given by $\left\langle c_{j}^{\dagger}c_{m}\right\rangle $
with $j,m\in X$. The R\'enyi entropies of subsystem $X$ satisfy
\citep{Peschel_2003}
\begin{equation}
S_{X}^{\left(n\right)}=\frac{1}{1-n}{\rm Tr}\ln\!\left[\left(C_{X}\!\right)^{n}+\left(\mathbb{I}-C_{X}\!\right)^{n}\right].\label{eq:Renyi-entropies-from-correlations}
\end{equation}
For the purpose of calculating R\'enyi entropies analytically, we
use the relation in Eq.~(\ref{eq:Renyi-entropies-from-correlations})
in two different ways. First, we may expand the logarithm in the form
of a power series, yielding
\begin{equation}
S_{X}^{\left(n\right)}=\frac{1}{1-n}\sum_{s=1}^{\infty}\frac{\left(-1\right)^{s+1}}{s}{\rm Tr}\!\left[\left\{ \left(C_{X}\!\right)^{n}+\left(\mathbb{I}-C_{X}\!\right)^{n}-\mathbb{I}\right\} ^{s}\right].\label{eq:Entropy-from-power-series}
\end{equation}
This expression implies that it suffices to produce a general formula
for the moments of the correlation matrix, i.e., to derive the asymptotics
of

\begin{align}
{\rm Tr}\!\left[\left(C_{X}\!\right)^{p}\right] & =\!\!\!\int_{\left[-\pi,\pi\right]^{p}}\!\!\frac{d^{p}k}{\left(2\pi\right)^{p}}\prod_{j=1}^{p}\!\widetilde{f}\!\left(k_{j}\right)\!\left[\sum_{m\in X}\!\!\left\langle m|k_{j-1}\right\rangle \!\left\langle k_{j}|m\right\rangle \right],\label{eq:Correlation-matrix-moment-general}
\end{align}
for any positive integer $p$ (we identify $k_{0}=k_{p}$). In our
case of interest, this indeed can be done in principle through the
stationary phase approximation (SPA) \citep{doi:10.1137/1.9780898719260},
on which we elaborate in \foreignlanguage{english}{Subsec.~\ref{subsec:Asymptotics-of-entropies-derivation}}.

The second way in which we may use Eq.~(\ref{eq:Renyi-entropies-from-correlations})
is through the root decomposition of the polynomial $z^{n}+\left(1-z\right)^{n}$,
which leads to the following expression:
\begin{equation}
S_{X}^{\left(n\right)}=\frac{1}{1-n}\sum_{\gamma=-\frac{n-1}{2}}^{\frac{n-1}{2}}\ln\det\!\left[\mathbb{I}+\left(e^{2\pi i\gamma/n}-1\right)C_{X}\!\right].\label{eq:Entropy-from-determinants}
\end{equation}
Then, the asymptotic analysis of the entropies can be carried out
if one has access to the asymptotics of the determinants in Eq.~(\ref{eq:Entropy-from-determinants}),
which is in fact true if $C_{X}$ has a Toeplitz (or block-Toeplitz)
structure, since in such a case the Szeg\H{o}-Widom asymptotic formula
can be employed \citep{WIDOM1974284}. As we explain below, this required
structure arises in our computation even when the restricted correlation
matrices of the subsystems of interest do not possess it themselves.

The treatment of R\'enyi negativities is closely analogous to that
of R\'enyi entropies. For a fermionic Gaussian state, negativities
between two subsystems $X_{1}$ and $X_{2}$ can be expressed using
$C_{X_{1}\cup X_{2}}$ and a transformed two-point correlation matrix
given by\foreignlanguage{english}{
\begin{equation}
C_{\Xi}=\frac{1}{2}\left[\mathbb{I}-\left(\mathbb{I}+\Gamma_{+}\Gamma_{-}\right)^{-1}\left(\Gamma_{+}+\Gamma_{-}\right)\right],\label{eq:Transformed-correlation-matrix}
\end{equation}
where we defined the covariance matrices
\begin{equation}
\Gamma_{\pm}=\left(\!\begin{array}{cc}
\pm i\mathbb{I}_{\left|X_{1}\right|} & 0\\
0 & \mathbb{I}_{\left|X_{2}\right|}
\end{array}\!\right)\!\left(\mathbb{I}-2C_{X_{1}\cup X_{2}}\right)\!\left(\!\begin{array}{cc}
\pm i\mathbb{I}_{\left|X_{1}\right|} & 0\\
0 & \mathbb{I}_{\left|X_{2}\right|}
\end{array}\!\right),
\end{equation}
assuming that the first $\left|X_{1}\right|$ indices designating
the entries of $C_{X_{1}\cup X_{2}}$ correspond to the sites of subsystem
$X_{1}$.} Using the matrix defined in Eq.~(\ref{eq:Transformed-correlation-matrix}),
R\'enyi negativities can be written as \citep{PhysRevB.95.165101,PhysRevB.97.165123,Shapourian_2019}
\begin{align}
{\cal E}^{\left(n\right)} & ={\rm Tr}\ln\!\left[\left(C_{\Xi}\right)^{n/2}+\left(\mathbb{I}-C_{\Xi}\right)^{n/2}\right]\nonumber \\
 & +\frac{n}{2}{\rm Tr}\ln\!\left[\left(C_{X_{1}\cup X_{2}}\right)^{2}+\left(\mathbb{I}-C_{X_{1}\cup X_{2}}\right)^{2}\right].\label{eq:Renyi-negativity-from-correlations-original}
\end{align}
Additionally, in Ref.~\citep{fraenkel2022extensive} we showed that
an equivalent way of writing this equality is
\begin{equation}
{\cal E}^{\left(n\right)}={\rm Tr}\ln\!\left[\prod_{\gamma=-\frac{n-1}{2}}^{\frac{n-1}{2}}\left(\mathbb{I}-C_{\gamma}\right)\right],\label{eq:Renyi-negativity-from-correlations}
\end{equation}
where we introduced for \foreignlanguage{english}{$\gamma=-\frac{n-1}{2},-\frac{n-3}{2},\ldots,\frac{n-1}{2}$}
the notation
\begin{equation}
C_{\gamma}=\left(\!\begin{array}{cc}
\left(1-e^{\frac{2\pi i\gamma}{n}}\right)\!\mathbb{I}_{\left|X_{1}\right|} & 0\\
0 & \left(1+e^{\frac{-2\pi i\gamma}{n}}\right)\!\mathbb{I}_{\left|X_{2}\right|}
\end{array}\!\right)C_{X_{1}\cup X_{2}}.\label{eq:Modified-correlation-matrix}
\end{equation}

\begin{widetext}As in the case of the R\'enyi entropies, the relation
to the correlation matrix in Eq.~(\ref{eq:Renyi-negativity-from-correlations})
can be used within two different approaches to access the asymptotics
of R\'enyi negatvities. In the first approach, we consider the series
expansion of Eq.~(\ref{eq:Renyi-negativity-from-correlations}),
\begin{equation}
{\cal E}^{\left(n\right)}=\sum_{s=1}^{\infty}\frac{\left(-1\right)^{s+1}}{s}{\rm Tr}\!\left[\left\{ \prod_{\gamma=-\frac{n-1}{2}}^{\frac{n-1}{2}}\left(\mathbb{I}-C_{\gamma}\right)-\mathbb{I}\right\} ^{\!s}\,\right],\label{eq:Negativity-from-power-series}
\end{equation}
and observe that it reduces the computation to that of an arbitrary
joint moment of the modified correlation matrices $C_{\gamma}$, that
is, a joint moment of the form
\begin{equation}
{\rm Tr}\!\left[C_{\gamma_{1}}\ldots C_{\gamma_{p}}\right]=\int_{\left[-\pi,\pi\right]^{p}}\!\!\frac{d^{p}k}{\left(2\pi\right)^{p}}\prod_{j=1}^{p}\widetilde{f}\!\left(k_{j}\right)\!\left[\left(1-e^{\frac{2\pi i\gamma_{j}}{n}}\right)\!\!\sum_{m\in X_{1}}\!\!\left\langle m|k_{j-1}\right\rangle \left\langle k_{j}|m\right\rangle +\left(1+e^{\frac{-2\pi i\gamma_{j}}{n}}\right)\!\!\sum_{m\in X_{2}}\!\!\left\langle m|k_{j-1}\right\rangle \left\langle k_{j}|m\right\rangle \right],\label{eq:Joint-correlation-matrix-moment-general}
\end{equation}
with $p$ being a positive integer. Eq.~(\ref{eq:Joint-correlation-matrix-moment-general})
bears clear resemblance to Eq.~(\ref{eq:Correlation-matrix-moment-general}),
and similarly its asymptotic form can be estimated in our case of
interest using the SPA (see details in Subsec.~\ref{subsec:Asymptotics-of-negativities-derivation}).
The second approach, which is analogous to the one that uses Eq.~(\ref{eq:Entropy-from-determinants})
in the case of the R\'enyi entropies, simply rewrites Eq.~(\ref{eq:Renyi-negativity-from-correlations})
as
\begin{equation}
{\cal E}^{\left(n\right)}=\sum_{\gamma=-\frac{n-1}{2}}^{\frac{n-1}{2}}\ln\det\!\left[\mathbb{I}-C_{\gamma}\right],\label{eq:Negativity-from-determinants}
\end{equation}
where again the asymptotics of the determinants can be analytically
calculated provided a certain structure of the modified correlation
matrices $C_{\gamma}$.

Lastly, we mention that, for a Gaussian state, the PRMI between $X_{1}$
and $X_{2}$ (defined in Eq.~(\ref{eq:PRMI-definition})) can also
be efficiently expressed using restricted two-point correlation matrices.
Indeed, the PRMI can be written as \citep{PhysRevLett.130.021603}
\begin{align}
{\cal D}^{\left(n\right)} & =\frac{1}{n-1}{\rm Tr}\ln\!\left[\left(C_{X_{1}\cup X_{2}}\right)^{n}\left(C_{X_{1}}\oplus C_{X_{2}}\right)^{1-n}+\left(\mathbb{I}-C_{X_{1}\cup X_{2}}\right)^{n}\left(\mathbb{I}-C_{X_{1}}\oplus C_{X_{2}}\right)^{1-n}\right].\label{eq:PRMI-from-correlations}
\end{align}
Eq.~(\ref{eq:PRMI-from-correlations}) is somewhat similar in its
form to Eq.~(\ref{eq:Renyi-entropies-from-correlations}) for the
R\'enyi entropy and Eq.~(\ref{eq:Renyi-negativity-from-correlations-original})
for the R\'enyi negativity, and indeed all of these formulas provide
efficient ways to numerically calculate the quantities that they represent.
However, the fact that the matrices $C_{X_{1}\cup X_{2}}$ and $C_{X_{1}}\oplus C_{X_{2}}$
in general do not commute makes it difficult to bring Eq.~(\ref{eq:PRMI-from-correlations})
to a form akin to Eq.~(\ref{eq:Entropy-from-determinants}) or Eq.~(\ref{eq:Negativity-from-determinants}),
and thus hinders the direct use of the same asymptotic techniques.
\selectlanguage{english}%

\subsubsection{Long-range limit of two-point correlations\label{subsec:Long-range-limit-correlations}}

\selectlanguage{american}%
Since we focus on sites outside of the impurity region (such that
$\left|j\right|,\left|m\right|>m_{0}$), Eqs.~(\ref{eq:Left-scattering-states})
and (\ref{eq:Right-scattering-states}) can be used to write down
the matrix element in Eq.~(\ref{eq:Correlation-function}) explicitly.
Furthermore, this explicit expression becomes simpler when considering
its long-range limit, i.e., when taking $d_{i}/\ell_{i}\to\infty$
with $d_{{\scriptscriptstyle L}}-d_{{\scriptscriptstyle R}}$ kept
fixed. As explained in Ref.~\citep{fraenkel2022extensive}, we may
use the Riemann-Lebesgue lemma to omit contributions to the matrix
element that vanish in this limit, since they correspond to Fourier
components with a diverging index. The two-point correlation function
is then given by\foreignlanguage{english}{
\begin{equation}
\left\langle c_{j}^{\dagger}c_{m}\right\rangle \longrightarrow\begin{cases}
\int_{-\pi}^{0}\frac{dk}{2\pi}\,\widetilde{f}\!\left(k\right)e^{-i\left(j-m\right)k}+\int_{0}^{\pi}\frac{dk}{2\pi}\left[\widetilde{f}\!\left(k\right){\cal T}\!\left(k\right)+\widetilde{f}\!\left(-k\right){\cal R}\!\left(k\right)\right]e^{-i\left(j-m\right)k} & j,m\in A_{{\scriptscriptstyle R}},\\
\int_{-\pi}^{0}\frac{dk}{2\pi}\,\widetilde{f}\!\left(-k\right)e^{i\left(j-m\right)k}+\int_{0}^{\pi}\frac{dk}{2\pi}\left[\widetilde{f}\!\left(-k\right){\cal T}\!\left(k\right)+\widetilde{f}\!\left(k\right){\cal R}\!\left(k\right)\right]e^{i\left(j-m\right)k} & j,m\in A_{{\scriptscriptstyle L}},\\
\int_{0}^{\pi}\frac{dk}{2\pi}\left[\widetilde{f}\!\left(k\right)t_{{\scriptscriptstyle L}}^{*}\!\left(k\right)r_{{\scriptscriptstyle L}}\!\left(k\right)+\widetilde{f}\!\left(-k\right)t_{{\scriptscriptstyle R}}\!\left(k\right)r_{{\scriptscriptstyle R}}^{*}\!\left(k\right)\right]e^{-i\left(j+m\right)k} & m\in A_{{\scriptscriptstyle L}}\text{ and }j\in A_{{\scriptscriptstyle R}},\\
\int_{0}^{\pi}\frac{dk}{2\pi}\left[\widetilde{f}\!\left(k\right)t_{{\scriptscriptstyle L}}\!\left(k\right)r_{{\scriptscriptstyle L}}^{*}\!\left(k\right)+\widetilde{f}\!\left(-k\right)t_{{\scriptscriptstyle R}}^{*}\!\left(k\right)r_{{\scriptscriptstyle R}}\!\left(k\right)\right]e^{i\left(j+m\right)k} & j\in A_{{\scriptscriptstyle L}}\text{ and }m\in A_{{\scriptscriptstyle R}}.
\end{cases}\label{eq:Correlation-function-long-range}
\end{equation}
\end{widetext}}

\selectlanguage{english}%
The analytical results presented in Sec.~\ref{sec:Results} depend
on \foreignlanguage{american}{$d_{{\scriptscriptstyle L}}$ and $d_{{\scriptscriptstyle R}}$}
only through \foreignlanguage{american}{$d_{{\scriptscriptstyle L}}-d_{{\scriptscriptstyle R}}$,
and in particular} hold in the long-range limit, a fact which arguably
constitutes the most remarkable aspect of our results. Accordingly,
the numerical results to which we compared them in Sec.~\ref{sec:Results}
(see Figs.~\ref{fig:Results-temperature-bias}, \ref{fig:Results-potential-bias},
and \ref{fig:Results-PRMI}) were computed using two-point correlation
matrices with their entries taken to the long-range limit given in
Eq.~(\ref{eq:Correlation-function-long-range}). The exact diagonalization
of the appropriate restricted correlation matrices allows to directly
use the $n\to1$ limits of Eqs.~(\ref{eq:Renyi-entropies-from-correlations})
and (\ref{eq:Renyi-negativity-from-correlations-original}), as well
as Eq.~(\ref{eq:PRMI-from-correlations}), to numerically compute
entanglement entropies, the fermionic negativity and the PRMI. Moreover,
the simplified correlation structure of Eq.~(\ref{eq:Correlation-function-long-range})
turns out to be essential for a central step in our analytical derivation,
even though this derivation begins from the full expression in Eq.~(\ref{eq:Correlation-function})
for the two-point correlation function; we elaborate on this point
in the following subsections.

\subsection{Asymptotics of R\'enyi entropies\label{subsec:Asymptotics-of-entropies-derivation}}

\subsubsection{Entropies of $A_{{\scriptscriptstyle L}}$ and $A_{{\scriptscriptstyle R}}$}

In deriving the asymptotic form of the R\'enyi entropies for the
two intervals $A_{{\scriptscriptstyle L}}$ and $A_{{\scriptscriptstyle R}}$,
we will rely on their relation with correlation matrix moments, which
was introduced in Subsec.~\ref{subsec:Two-point-correlations}. Following
the same methodology as the one that guided the analysis in Ref.~\citep{fraenkel2022extensive},
we substitute the eigenstate wavefunctions in Eqs.~(\ref{eq:Left-scattering-states})
and (\ref{eq:Right-scattering-states}) into the expression for the
correlation matrix moment given in Eq.~(\ref{eq:Correlation-matrix-moment-general}).
We employ the identity
\begin{align}
\sum_{m=r+1}^{r+s}\exp\!\left[im\left(k_{j-1}-k_{j}\right)\right] & =s\,{\cal W}_{r}\!\left(\frac{k_{j-1}-k_{j}}{2}\right)\nonumber \\
 & \times\int_{0}^{1}d\xi\exp\!\left[is\left(k_{j-1}-k_{j}\right)\xi\right],\label{eq:Sum-as-integral-identity}
\end{align}
where we defined ${\cal W}_{r}\!\left(x\right)=\frac{1}{\sin x}xe^{i\left(2r+1\right)x}$,
in order to bring the $p$th correlation matrix moment to the general
form
\begin{equation}
{\rm Tr}\!\left[\left(C_{A_{i}}\right)^{p}\right]=\sum_{\overrightarrow{\tau},\overrightarrow{\sigma}\in\left\{ 0,1\right\} ^{\otimes p}}{\cal G}\!\left(\overrightarrow{\tau},\overrightarrow{\sigma}\right),\label{eq:Correlation-matrix-moment-SPA-form}
\end{equation}
with

\begin{align}
{\cal G}\!\left(\overrightarrow{\tau},\overrightarrow{\sigma}\right) & ={\ell_{i}}^{p}\int_{\left[-\pi,\pi\right]^{p}}\frac{d^{p}k}{\left(2\pi\right)^{p}}\int_{\left[0,1\right]^{p}}d^{p}\xi\,g_{\overrightarrow{\tau}\!,\!\overrightarrow{\sigma}}\!\left(\overrightarrow{k}\right)\nonumber \\
 & \times\exp\!\left[i\ell_{i}\sum_{j=1}^{p}\left(\left(-1\right)^{\tau_{j-1}}\!k_{j-1}-\left(-1\right)^{\sigma_{j}}\!k_{j}\right)\xi_{j}\right].\label{eq:Correlation-matrix-moment-SPA-form-summands}
\end{align}
The particular forms of the functions $g_{\overrightarrow{\tau}\!,\!\overrightarrow{\sigma}}$
depend on the scattering matrix associated with the impurity, as well
as on the occupation factor $\widetilde{f}\!\left(k\right)$ associated
with the reservoirs. Importantly, however, these functions are independent
of $\ell_{i}$.

In Ref.~\citep{fraenkel2022extensive} we performed the asymptotic
analysis of moments that were brought to the same form as in Eq.~(\ref{eq:Correlation-matrix-moment-SPA-form}).
This was done by applying to the integrals in Eq.~(\ref{eq:Correlation-matrix-moment-SPA-form-summands})
the SPA, according to which leading-order contributions arise from
stationary points of the function appearing inside the exponent. There,
we showed that in Eq.~(\ref{eq:Correlation-matrix-moment-SPA-form})
only the terms with $\overrightarrow{\tau}=\overrightarrow{\sigma}$
may have a leading-order ${\cal O}\!\left(\ell_{i}\right)$ contribution
to the moment, whereas terms with $\overrightarrow{\tau}\neq\overrightarrow{\sigma}$
have an ${\cal O}\!\left({\ell_{i}}^{0}\right)$ contribution at most.
The leading-order asymptotics is furthermore given by the formula
\begin{align}
 & {\rm Tr}\!\left[\left(C_{A_{i}}\right)^{p}\right]\nonumber \\
 & \sim\ell_{i}\!\!\int_{-\pi}^{\pi}\!\!\frac{dk}{2\pi}\!\sum_{\overrightarrow{\sigma}\in\left\{ 0,1\right\} ^{\otimes p}}\!\!\!\!g_{\overrightarrow{\sigma}\!,\!\overrightarrow{\sigma}}\!\!\left(k,\left(-1\right)^{\sigma_{1}+\sigma_{2}}\!k,\ldots,\left(-1\right)^{\sigma_{1}+\sigma_{p}}\!k\right),\label{eq:Correlation-matrix-moment-SPA-asymptotics}
\end{align}
which we can then express more explicitly using the particular functions
$g_{\overrightarrow{\tau}\!,\!\overrightarrow{\sigma}}$ that satisfy
the equality in Eq.~(\ref{eq:Correlation-matrix-moment-SPA-form}).
In Appendix \ref{sec:Moments-for-single-intervals-appendix} we go
through the details of the computation, arriving at the following
asymptotic formulas:
\begin{align}
{\rm Tr}\!\left[\left(C_{A_{L}}\right)^{p}\right] & \sim\ell_{{\scriptscriptstyle L}}\bigg[\int_{0}^{\pi}\frac{dk}{2\pi}\left(f_{{\scriptscriptstyle L}}\!\left(k\right)\right)^{p}\nonumber \\
 & +\int_{0}^{\pi}\frac{dk}{2\pi}\left({\cal R}\!\left(k\right)f_{{\scriptscriptstyle L}}\!\left(k\right)+{\cal T}\!\left(k\right)f_{{\scriptscriptstyle R}}\!\left(k\right)\right)^{p}\bigg],\nonumber \\
{\rm Tr}\!\left[\left(C_{A_{R}}\right)^{p}\right] & \sim\ell_{{\scriptscriptstyle R}}\bigg[\int_{0}^{\pi}\frac{dk}{2\pi}\left(f_{{\scriptscriptstyle R}}\!\left(k\right)\right)^{p}\nonumber \\
 & +\int_{0}^{\pi}\frac{dk}{2\pi}\left({\cal T}\!\left(k\right)f_{{\scriptscriptstyle L}}\!\left(k\right)+{\cal R}\!\left(k\right)f_{{\scriptscriptstyle R}}\!\left(k\right)\right)^{p}\bigg].\label{eq:Correlation-matrix-moment-interval-asymptotics}
\end{align}
Finally, by substituting Eq.~(\ref{eq:Correlation-matrix-moment-interval-asymptotics})
into the power series expansion in Eq.~(\ref{eq:Entropy-from-power-series}),
we obtain the following expressions for the R\'enyi entropies of
$A_{{\scriptscriptstyle L}}$ and $A_{{\scriptscriptstyle R}}$:
\begin{align}
S_{A_{L}}^{\left(n\right)} & \sim\frac{\ell_{{\scriptscriptstyle L}}}{1-n}\int_{0}^{\pi}\frac{dk}{2\pi}\bigg\{\ln\!\left[\left(f_{{\scriptscriptstyle L}}\right)^{n}+\left(1-f_{{\scriptscriptstyle L}}\right)^{n}\right]\nonumber \\
 & +\ln\!\left[\left({\cal R}f_{{\scriptscriptstyle L}}+{\cal T}f_{{\scriptscriptstyle R}}\right)^{n}+\left(1-{\cal R}f_{{\scriptscriptstyle L}}-{\cal T}f_{{\scriptscriptstyle R}}\right)^{n}\right]\bigg\},\nonumber \\
S_{A_{R}}^{\left(n\right)} & \sim\frac{\ell_{{\scriptscriptstyle R}}}{1-n}\int_{0}^{\pi}\frac{dk}{2\pi}\bigg\{\ln\!\left[\left(f_{{\scriptscriptstyle R}}\right)^{n}+\left(1-f_{{\scriptscriptstyle R}}\right)^{n}\right]\nonumber \\
 & +\ln\!\left[\left({\cal T}f_{{\scriptscriptstyle L}}+{\cal R}f_{{\scriptscriptstyle R}}\right)^{n}+\left(1-{\cal T}f_{{\scriptscriptstyle L}}-{\cal R}f_{{\scriptscriptstyle R}}\right)^{n}\right]\bigg\}.\label{eq:Interval-Renyi-entropy-asymptotics}
\end{align}

\subsubsection{Entropies of $A_{{\scriptscriptstyle L}}\cup A_{{\scriptscriptstyle R}}$}

For convenience, we introduce here the notations $A=A_{{\scriptscriptstyle L}}\cup A_{{\scriptscriptstyle R}}$
and $\Delta\ell_{i}=\ell_{i}-\ell_{{\rm mirror}}$ ($i={\scriptstyle L,R}$).
Similarly to the case of the intervals $A_{{\scriptscriptstyle L}}$
and $A_{{\scriptscriptstyle R}}$, the R\'enyi entropies of their
union $A$ can be computed by using the SPA to produce a general asymptotic
expression for an integer moment of $C_{A}$, and by subsequently
summing the power series in Eq.~(\ref{eq:Entropy-from-power-series}).
The application of the SPA in this case, however, is considerably
more intricate technically. Instead, we choose to rely on the SPA
only to prove a simple relation between moments of $C_{A}$ and moments
of other restricted correlation matrices, a relation that leads to
a useful decomposition of R\'enyi entropies of $A$.

In Appendix \ref{sec:Moments-for-two-intervals-appendix} we employ
the SPA (essentially repeating an argument that we originally introduced
in Ref.~\citep{fraenkel2022extensive}) to show that, to a leading
order, the moments of $C_{A}$ can be decomposed into a sum of independent
contributions in the form
\begin{align}
{\rm Tr}\!\left[\left(C_{A}\right)^{p}\right] & \sim\frac{\Delta\ell_{{\scriptscriptstyle L}}}{\ell_{{\scriptscriptstyle L}}}{\rm Tr}\!\left[\left(C_{A_{L}}\right)^{p}\right]+\frac{\Delta\ell_{{\scriptscriptstyle R}}}{\ell_{{\scriptscriptstyle R}}}{\rm Tr}\!\left[\left(C_{A_{R}}\right)^{p}\right]\nonumber \\
 & +{\rm Tr}\!\left[\left(C_{{\rm mirror}}\right)^{p}\right],\label{eq:Correlation-matrix-moment-combined-decomposition}
\end{align}
where $C_{{\rm mirror}}$ is the restricted two-point correlation
matrix of the subsystem $A_{{\rm mirror}}=\left(A_{{\scriptscriptstyle L}}\cap\bar{A}_{{\scriptscriptstyle R}}\right)\cup\left(\bar{A}_{{\scriptscriptstyle L}}\cap A_{{\scriptscriptstyle R}}\right)$,
i.e., the subsystem containing all sites in $A_{{\scriptscriptstyle L}}$
that have mirroring sites in $A_{{\scriptscriptstyle R}}$ and vice
versa. For the first two moments appearing in the right-hand side
of Eq.~(\ref{eq:Correlation-matrix-moment-combined-decomposition})
we can use the result of Eq.~(\ref{eq:Correlation-matrix-moment-interval-asymptotics}),
while a separate treatment is required for the moment of $C_{{\rm mirror}}$.

The matrix $C_{{\rm mirror}}$ is of size $2\ell_{{\rm mirror}}\times2\ell_{{\rm mirror}}$,
and, as already mentioned, its moment can be estimated asymptotically
for $\ell_{{\rm mirror}}\gg1$ using the SPA, as was indeed done in
Ref.~\citep{fraenkel2022extensive} for the zero-temperature case.
There is, however, an alternative approach that simplifies the calculation.
The SPA analysis presented in Appendix \ref{sec:Moments-for-two-intervals-appendix}
shows that, to a leading order, the moment of $C_{{\rm mirror}}$
is independent of $\max\!\left\{ d_{{\scriptscriptstyle L}},d_{{\scriptscriptstyle R}}\right\} $,
which determines the absolute distance of the mirroring sites from
the impurity region. This implies that we may take the limit $d_{i}\to\infty$
(with $d_{{\scriptscriptstyle L}}-d_{{\scriptscriptstyle R}}$ kept
fixed) and the leading-order value of the moment will remain the same.
Since this is true for any moment of $C_{{\rm mirror}}$, it is true
(by virtue of Eq.~(\ref{eq:Entropy-from-power-series})) also for
the R\'enyi entropies of $A_{{\rm mirror}}$. In other words, Eqs.~(\ref{eq:Entropy-from-power-series})
and (\ref{eq:Correlation-matrix-moment-combined-decomposition}) allow
us to write
\begin{equation}
S_{A}^{\left(n\right)}\sim\frac{\Delta\ell_{{\scriptscriptstyle L}}}{\ell_{{\scriptscriptstyle L}}}S_{A_{L}}^{\left(n\right)}+\frac{\Delta\ell_{{\scriptscriptstyle R}}}{\ell_{{\scriptscriptstyle R}}}S_{A_{R}}^{\left(n\right)}+S_{{\rm mirror}}^{\left(n\right)},\label{eq:Entropy-combined-decomposition}
\end{equation}
with $S_{{\rm mirror}}^{\left(n\right)}$ designating the $n$th R\'enyi
entropy of $A_{{\rm mirror}}$, and to regard $A_{{\rm mirror}}$
as the union of two mirroring intervals that are infinitely far apart.

Following this reasoning, we can invoke the long-range limit of the
two-point correlation function given in Eq.~(\ref{eq:Correlation-function-long-range}),
which leads to a simplified structure of $C_{{\rm mirror}}$. Expressing
the relation between $S_{{\rm mirror}}^{\left(n\right)}$ and $C_{{\rm mirror}}$
using Eq.~(\ref{eq:Entropy-from-determinants}), this simplified
structure enables the use of a powerful asymptotic technique. Indeed,
using Eq\@.~(\ref{eq:Correlation-function-long-range}) we can arrange
the long-range limit of $C_{{\rm mirror}}$ in the form of $\ell_{{\rm mirror}}\times\ell_{{\rm mirror}}$
blocks, each of size $2\times2$, with the entries of the block $\Phi^{\left(j,m\right)}$
($j,m=1,\ldots,\ell_{{\rm mirror}}$) depending only on the difference
$j-m$; in that form, $C_{{\rm mirror}}$ is a block-Toeplitz matrix.
The explicit form of the block $\Phi^{\left(j,m\right)}$ can be conveniently
written as
\begin{equation}
\Phi^{\left(j,m\right)}=\int_{-\pi}^{\pi}\frac{dk}{2\pi}\Phi\!\left(k\right)e^{-i\left(j-m\right)k},\label{eq:Block-Toeplitz-mirror}
\end{equation}
where the entries of the $2\times2$ block symbol $\Phi\!\left(k\right)$
are given by
\begin{align}
\Phi_{11} & =\begin{cases}
\widetilde{f}\!\left(-k\right) & -\pi<k<0,\\
\widetilde{f}\!\left(-k\right){\cal T}\!\left(k\right)+\widetilde{f}\!\left(k\right){\cal R}\!\left(k\right) & 0<k<\pi,
\end{cases}\nonumber \\
\Phi_{22} & =\begin{cases}
\widetilde{f}\!\left(k\right) & -\pi<k<0,\\
\widetilde{f}\!\left(k\right){\cal T}\!\left(k\right)+\widetilde{f}\!\left(-k\right){\cal R}\!\left(k\right) & 0<k<\pi,
\end{cases}\nonumber \\
\Phi_{12} & =\begin{cases}
0 & -\pi<k<0,\\
\left(\widetilde{f}\!\left(k\right)-\widetilde{f}\!\left(-k\right)\right)t_{{\scriptscriptstyle L}}\!\left(k\right)r_{{\scriptscriptstyle L}}^{*}\!\left(k\right) & 0<k<\pi,
\end{cases}\label{eq:Block-Toeplitz-symbol-mirror}
\end{align}
and $\Phi_{21}=\Phi_{12}^{*}$. Note that we used the fact that $t_{{\scriptscriptstyle R}}^{*}r_{{\scriptscriptstyle R}}=-t_{{\scriptscriptstyle L}}r_{{\scriptscriptstyle L}}^{*}$
due to the unitarity of the scattering matrix.

Next, we observe that in each determinant featured in Eq.~(\ref{eq:Entropy-from-determinants})
appears a block-Toeplitz matrix with the block symbol $\mathbb{I}_{2}+\left(e^{2\pi i\gamma/n}-1\right)\Phi$.
The Szeg\H{o}-Widom formula \citep{WIDOM1974284} states that, given
a block-Toeplitz matrix $T_{\ell}\!\left[\Psi\right]$ defined by
a continuous block symbol $\Psi\!\left(k\right)$ and comprised of
$\ell\times\ell$ blocks, the leading-order large-$\ell$ asymptotics
of its determinant is given by
\begin{equation}
\ln\det T_{\ell}\!\left[\Psi\right]\sim\ell\int_{-\pi}^{\pi}\frac{dk}{2\pi}\ln\det\Psi\!\left(k\right).\label{eq:Szego-Widom formula}
\end{equation}
Accordingly, we find that
\begin{align}
 & \ln\det\!\left[\mathbb{I}+\left(e^{2\pi i\gamma/n}-1\right)C_{{\rm mirror}}\right]\nonumber \\
 & \sim\ell_{{\rm mirror}}\int_{-\pi}^{\pi}\frac{dk}{2\pi}\ln\det\!\left[\mathbb{I}_{2}+\left(e^{2\pi i\gamma/n}-1\right)\Phi\!\left(k\right)\right]\nonumber \\
 & =\ell_{{\rm mirror}}\int_{-\pi}^{\pi}\frac{dk}{\pi}\ln\!\left[1+\left(e^{2\pi i\gamma/n}-1\right)\widetilde{f}\!\left(k\right)\right],
\end{align}
and then, by summing over the index $\gamma$ in Eq.~(\ref{eq:Entropy-from-determinants}),
we obtain
\begin{equation}
S_{{\rm mirror}}^{\left(n\right)}\sim\frac{\ell_{{\rm mirror}}}{1-n}\int_{-\pi}^{\pi}\frac{dk}{\pi}\ln\!\left[\left(\widetilde{f}\!\left(k\right)\right)^{n}+\left(1-\widetilde{f}\!\left(k\right)\right)^{n}\right].\label{eq:Mirror-Renyi-entropy-asymptotics}
\end{equation}
Finally, combining Eqs. (\ref{eq:Interval-Renyi-entropy-asymptotics}),
(\ref{eq:Entropy-combined-decomposition}) and (\ref{eq:Mirror-Renyi-entropy-asymptotics}),
we may write the R\'enyi entropies of subsystem $A$ as
\begin{align}
S_{A}^{\left(n\right)} & \sim\frac{\ell_{{\scriptscriptstyle L}}+\ell_{{\rm mirror}}}{1-n}\!\int_{0}^{\pi}\!\!\frac{dk}{2\pi}\ln\!\left[\left(f_{{\scriptscriptstyle L}}\right)^{n}+\left(1\!-\!f_{{\scriptscriptstyle L}}\right)^{n}\right]\nonumber \\
 & +\frac{\ell_{{\scriptscriptstyle R}}+\ell_{{\rm mirror}}}{1-n}\!\int_{0}^{\pi}\!\!\frac{dk}{2\pi}\ln\!\left[\left(f_{{\scriptscriptstyle R}}\right)^{n}+\left(1\!-\!f_{{\scriptscriptstyle R}}\right)^{n}\right]\nonumber \\
 & +\frac{\Delta\ell_{{\scriptscriptstyle L}}}{1-n}\!\int_{0}^{\pi}\!\!\frac{dk}{2\pi}\ln\!\left[\left({\cal R}f_{{\scriptscriptstyle L}}\!+\!{\cal T}f_{{\scriptscriptstyle R}}\right)^{n}+\left(1\!-\!{\cal R}f_{{\scriptscriptstyle L}}\!-\!{\cal T}f_{{\scriptscriptstyle R}}\right)^{n}\right]\nonumber \\
 & +\frac{\Delta\ell_{{\scriptscriptstyle R}}}{1-n}\!\int_{0}^{\pi}\!\!\frac{dk}{2\pi}\ln\!\left[\left({\cal T}f_{{\scriptscriptstyle L}}\!+\!{\cal R}f_{{\scriptscriptstyle R}}\right)^{n}+\left(1\!-\!{\cal T}f_{{\scriptscriptstyle L}}\!-\!{\cal R}f_{{\scriptscriptstyle R}}\right)^{n}\right].\label{eq:Combined-Renyi-entropy-asymptotics}
\end{align}
Eqs.~(\ref{eq:Interval-Renyi-entropy-asymptotics}) and (\ref{eq:Combined-Renyi-entropy-asymptotics}),
when combined, lead to the expression in Eq.~(\ref{eq:Renyi-MI-main-result})
for the RMI.

We conclude this part by noting that the approach we took to compute
the entropies of $A$ can be similarly used for the computation of
the entropies of $A_{{\scriptscriptstyle L}}$ and $A_{{\scriptscriptstyle R}}$.
That is, instead of using the SPA throughout the entire computation
-- estimating exactly the correlation matrix moments and then summing
the power series in Eq.~(\ref{eq:Entropy-from-power-series}) --
we can use the SPA only to show that $S_{A_{i}}^{\left(n\right)}$
is independent of $d_{i}$ to a leading order. Then, we can rely on
this fact to compute $S_{A_{i}}^{\left(n\right)}$ through Eq.~(\ref{eq:Entropy-from-determinants})
by substituting the long-range limit of $C_{A_{i}}$, which is of
a Toeplitz form, as can be checked via Eq.~(\ref{eq:Correlation-function-long-range}).
The Szeg\H{o}-Widom formula can therefore be used to complete the
calculation. Nevertheless, the direct application of the SPA to the
computation of correlation matrix moments is relatively simple in
the case of $A_{{\scriptscriptstyle L}}$ and $A_{{\scriptscriptstyle R}}$,
and we chose to present these two different variations of our analytical
methodology in order to emphasize its versatility.

\subsection{Asymptotics of R\'enyi negativities\label{subsec:Asymptotics-of-negativities-derivation}}

Here we address the calculation of the R\'enyi negativity ${\cal E}^{\left(n\right)}$,
which yields upon analytic continuation the asymptotics of the negativity,
reported in Eq.~(\ref{eq:Negativity-main-result}). The derivation
of ${\cal E}^{\left(n\right)}$ largely parallels that of $S_{A}^{\left(n\right)}$
($A=A_{{\scriptscriptstyle L}}\cup A_{{\scriptscriptstyle R}}$) discussed
in Subsec.~\ref{subsec:Asymptotics-of-entropies-derivation}. We
will again use the notation $\Delta\ell_{i}=\ell_{i}-\ell_{{\rm mirror}}$
($i={\scriptstyle L,R}$).

In Subsec.~\ref{subsec:Two-point-correlations} we explained that
the R\'enyi negativity ${\cal E}^{\left(n\right)}$ can be expressed
as a series of joint moments of the form given in Eq.~(\ref{eq:Joint-correlation-matrix-moment-general})
(with $X_{1}$ and $X_{2}$ standing for $A_{{\scriptscriptstyle L}}$
and $A_{{\scriptscriptstyle R}}$, respectively), such that by computing
a general expression for these joint moments we can obtain an expression
for ${\cal E}^{\left(n\right)}$. As we explain in Appendix \ref{sec:Moments-for-two-intervals-appendix},
a decomposition of the joint moments which is analogous to \foreignlanguage{american}{Eq.~(\ref{eq:Correlation-matrix-moment-combined-decomposition})
(used in the computation of $S_{A}^{\left(n\right)}$) is possible
here due to the same SPA argument.} Within the steady state of interest
and to a leading order, the joint moments satisfy\foreignlanguage{american}{
\begin{align}
{\rm Tr}\!\left[C_{\gamma_{1}}\ldots C_{\gamma_{p}}\right] & \sim\frac{\Delta\ell_{{\scriptscriptstyle L}}}{\ell_{{\scriptscriptstyle L}}}\left(\prod_{j=1}^{p}\left(1-e^{\frac{2\pi i\gamma_{j}}{n}}\right)\right){\rm Tr}\!\left[\left(C_{A_{L}}\right)^{p}\right]\nonumber \\
 & +\frac{\Delta\ell_{{\scriptscriptstyle R}}}{\ell_{{\scriptscriptstyle R}}}\left(\prod_{j=1}^{p}\left(1+e^{\frac{-2\pi i\gamma_{j}}{n}}\right)\right){\rm Tr}\!\left[\left(C_{A_{R}}\right)^{p}\right]\nonumber \\
 & +{\rm Tr}\!\left[C_{{\rm mirror},\gamma_{1}}\ldots C_{{\rm mirror},\gamma_{p}}\right],\label{eq:Joint-moment-SPA-decomposition}
\end{align}
where $C_{{\rm mirror},\gamma}$ is a modified version of $C_{{\rm mirror}}$
(the latter was defined right below Eq.~(\ref{eq:Correlation-matrix-moment-combined-decomposition})),
given by Eq.~(\ref{eq:Modified-correlation-matrix}) with $X_{1}=A_{{\scriptscriptstyle L}}\cap\bar{A}_{{\scriptscriptstyle R}}$
and $X_{2}=\bar{A}_{{\scriptscriptstyle L}}\cap A_{{\scriptscriptstyle R}}$.
Substituting Eq.~(\ref{eq:Joint-moment-SPA-decomposition}) into
the power series expansion of the R\'enyi negativity in Eq.~(\ref{eq:Negativity-from-power-series}),
we find upon summation that
\begin{equation}
{\cal E}^{\left(n\right)}\sim\left(1-n\right)\left[\frac{\Delta\ell_{{\scriptscriptstyle L}}}{\ell_{{\scriptscriptstyle L}}}S_{A_{L}}^{\left(n\right)}+\frac{\Delta\ell_{{\scriptscriptstyle R}}}{\ell_{{\scriptscriptstyle R}}}S_{A_{R}}^{\left(n\right)}\right]+{\cal E}_{{\rm mirror}}^{\left(n\right)},\label{eq:Negativity-SPA-decomposition}
\end{equation}
with ${\cal E}_{{\rm mirror}}^{\left(n\right)}$ standing for the
negativity between the two mirroring intervals comprising $A_{{\rm mirror}}$,
i.e., $A_{{\scriptscriptstyle L}}\cap\bar{A}_{{\scriptscriptstyle R}}$
and $\bar{A}_{{\scriptscriptstyle L}}\cap A_{{\scriptscriptstyle R}}$. }

\selectlanguage{american}%
Moreover, as with the moments of $C_{{\rm mirror}}$ (see the discussion
following Eq.~(\ref{eq:Correlation-matrix-moment-combined-decomposition})),
the argument laid out in Appendix \ref{sec:Moments-for-two-intervals-appendix}
implies that joint moments of the matrices $C_{{\rm mirror},\gamma}$
are independent of the distance between $A_{{\scriptscriptstyle L}}\cap\bar{A}_{{\scriptscriptstyle R}}$
and $\bar{A}_{{\scriptscriptstyle L}}\cap A_{{\scriptscriptstyle R}}$,
such that we may assume that this distance is much larger than $\ell_{{\rm mirror}}$
and use the long-range limit of the two-point correlation function
(given in Eq.~(\ref{eq:Correlation-function-long-range})) to express
the entries of $C_{{\rm mirror}}$.\foreignlanguage{english}{ Consequently,
the matrices }$C_{{\rm mirror},\gamma}$ can be written as block-Toeplitz
matrices with \foreignlanguage{english}{$\ell_{{\rm mirror}}\times\ell_{{\rm mirror}}$}
blocks of size $2\times2$ each. The associated block symbol is given
by
\begin{equation}
\Phi_{\gamma}\!\left(k\right)=\left(\!\begin{array}{cc}
1-e^{\frac{2\pi i\gamma}{n}} & 0\\
0 & 1+e^{\frac{-2\pi i\gamma}{n}}
\end{array}\!\right)\Phi\!\left(k\right),\label{eq:Block-Toeplitz-symbol-negativity}
\end{equation}
with $\Phi$ being the block symbol of $C_{{\rm mirror}}$, the entries
of which are given explicitly in Eq.~(\ref{eq:Block-Toeplitz-symbol-mirror}).

We may now exploit the block-Toeplitz structure of the matrices $C_{{\rm mirror},\gamma}$
to obtain the asymptotics of ${\cal E}_{{\rm mirror}}^{\left(n\right)}$,
given the relation in Eq.~(\ref{eq:Negativity-from-determinants})
between these matrices and the R\'enyi negativity. Employing the
Szeg\H{o}-Widom formula of Eq.~(\ref{eq:Szego-Widom formula}),
in Appendix \ref{sec:Renyi-negativity-appendix} we compute the leading-order
asymptotics of $\ln\det\!\left[\mathbb{I}-C_{{\rm mirror},\gamma}\right]$
and perform the summation over the index $\gamma$ in Eq.~(\ref{eq:Negativity-from-determinants}),
finding that
\begin{align}
{\cal E}_{{\rm mirror}}^{\left(n\right)} & \sim\ell_{{\rm mirror}}\int_{-\pi}^{\pi}\frac{dk}{2\pi}\ln\!\left[\left(\widetilde{f}\!\left(k\right)\right)^{n}+\left(1-\widetilde{f}\!\left(k\right)\right)^{n}\right]\nonumber \\
 & +\ell_{{\rm mirror}}\int_{0}^{\pi}\frac{dk}{2\pi}\ln{\cal Y}_{n}\!\left(k\right),\label{eq:Mirror-Renyi-negativity-asymptotics}
\end{align}
where we defined\begin{widetext}
\begin{align}
{\cal Y}_{n} & =\left[{\cal T}f_{{\scriptscriptstyle L}}+{\cal R}f_{{\scriptscriptstyle R}}-f_{{\scriptscriptstyle L}}f_{{\scriptscriptstyle R}}\right]^{n}+\left[{\cal R}f_{{\scriptscriptstyle L}}+{\cal T}f_{{\scriptscriptstyle R}}-f_{{\scriptscriptstyle L}}f_{{\scriptscriptstyle R}}\right]^{n}+\left[\sqrt{\left[\frac{1\!-\!f_{{\scriptscriptstyle L}}\!-\!f_{{\scriptscriptstyle R}}\!+\!2f_{{\scriptscriptstyle L}}f_{{\scriptscriptstyle R}}}{2}\right]^{2}+{\cal TR}\left(f_{{\scriptscriptstyle L}}\!-\!f_{{\scriptscriptstyle R}}\right)^{2}}+\frac{1\!-\!f_{{\scriptscriptstyle L}}\!-\!f_{{\scriptscriptstyle R}}}{2}\right]^{n}\nonumber \\
 & +\left[\sqrt{\left[\frac{1\!-\!f_{{\scriptscriptstyle L}}\!-\!f_{{\scriptscriptstyle R}}\!+\!2f_{{\scriptscriptstyle L}}f_{{\scriptscriptstyle R}}}{2}\right]^{2}+{\cal TR}\left(f_{{\scriptscriptstyle L}}\!-\!f_{{\scriptscriptstyle R}}\right)^{2}}-\frac{1\!-\!f_{{\scriptscriptstyle L}}\!-\!f_{{\scriptscriptstyle R}}}{2}\right]^{n}.\label{eq:Negativity-auxiliary-polynomial}
\end{align}
Finally, by combining Eqs.~(\ref{eq:Negativity-SPA-decomposition})
and (\ref{eq:Mirror-Renyi-negativity-asymptotics}), we arrive at
the desired expression for the R\'enyi negativity between \foreignlanguage{english}{$A_{{\scriptscriptstyle L}}$
and $A_{{\scriptscriptstyle R}}$}:
\begin{align}
{\cal E}^{\left(n\right)} & \sim\ell_{{\rm mirror}}\!\int_{0}^{\pi}\!\frac{dk}{2\pi}\ln{\cal Y}_{n}+\ell_{{\scriptscriptstyle L}}\!\int_{0}^{\pi}\!\frac{dk}{2\pi}\ln\!\left[\left(f_{{\scriptscriptstyle L}}\right)^{n}+\left(1-f_{{\scriptscriptstyle L}}\right)^{n}\right]+\ell_{{\scriptscriptstyle R}}\!\int_{0}^{\pi}\!\frac{dk}{2\pi}\ln\!\left[\left(f_{{\scriptscriptstyle R}}\right)^{n}+\left(1-f_{{\scriptscriptstyle R}}\right)^{n}\right]\nonumber \\
 & +\Delta\ell_{{\scriptscriptstyle L}}\!\int_{0}^{\pi}\!\frac{dk}{2\pi}\ln\!\left[\left({\cal R}f_{{\scriptscriptstyle L}}+{\cal T}f_{{\scriptscriptstyle R}}\right)^{n}+\left(1-{\cal R}f_{{\scriptscriptstyle L}}-{\cal T}f_{{\scriptscriptstyle R}}\right)^{n}\right]+\Delta\ell_{{\scriptscriptstyle R}}\!\int_{0}^{\pi}\!\frac{dk}{2\pi}\ln\!\left[\left({\cal T}f_{{\scriptscriptstyle L}}+{\cal R}f_{{\scriptscriptstyle R}}\right)^{n}+\left(1-{\cal T}f_{{\scriptscriptstyle L}}-{\cal R}f_{{\scriptscriptstyle R}}\right)^{n}\right].
\end{align}
\end{widetext}The fermionic negativity in Eq.~(\ref{eq:Negativity-main-result})
is obtained by taking the $n\to1$ limit of the above expression.
\selectlanguage{english}%

\subsection{Asymptotics of Petz-R\'enyi mutual information\label{subsec:Asymptotics-of-PRMI-derivation}}

In this part, we briefly summarize the calculation leading to Eq.~(\ref{eq:PRMI-main-result}),
which states the analytical form of the PRMI ${\cal D}^{\left(n\right)}$
between $A_{{\scriptscriptstyle L}}$ and $A_{{\scriptscriptstyle R}}$.
The derivation is more conjectural compared to those discussed in
Subsecs.~\ref{subsec:Asymptotics-of-entropies-derivation} and \ref{subsec:Asymptotics-of-negativities-derivation},
owing to the difficulty (which was already mentioned in Subsec.~\ref{subsec:Two-point-correlations})
in applying the same analytical techniques to the expression in Eq.~(\ref{eq:PRMI-from-correlations})
of the PRMI in terms of two-point correlation matrices. The validity
of the final result is corroborated by its comparison to numerical
results, presented in Fig.~\ref{fig:Results-PRMI} and discussed
at the end of Sec.~\ref{sec:Results}.

Like in the case of Eqs\@.~(\ref{eq:Entropy-from-power-series})
and (\ref{eq:Negativity-from-power-series}), Eq.~(\ref{eq:PRMI-from-correlations})
can be expanded to a power series, reducing the calculation of the
PRMI to that of joint moments of $C_{A}$ and $C_{A_{L}}\oplus C_{A_{R}}$.
By the same argument that is detailed in Appendix \ref{sec:Moments-for-two-intervals-appendix},
at the leading order these moments can be decomposed into independent
contributions from $A_{{\rm mirror}}$, $A_{{\scriptscriptstyle L}}\setminus\bar{A}_{{\scriptscriptstyle R}}$
and $A_{{\scriptscriptstyle R}}\setminus\bar{A}_{{\scriptscriptstyle L}}$;
since the matrices $C_{A}$ and $C_{A_{L}}\oplus C_{A_{R}}$ are identical
when projected onto either $A_{{\scriptscriptstyle L}}\setminus\bar{A}_{{\scriptscriptstyle R}}$
or $A_{{\scriptscriptstyle R}}\setminus\bar{A}_{{\scriptscriptstyle L}}$,
the contributions from these two subsystems fall off, leaving us with\foreignlanguage{american}{
\begin{align}
{\cal D}^{\left(n\right)} & \sim\frac{1}{n-1}\ln\det\!\bigg[\!\left(C_{{\rm mirror}}\right)^{n}\left(C_{A_{L}\cap\bar{A}_{R}}\oplus C_{\bar{A}_{L}\cap A_{R}}\right)^{1-n}\nonumber \\
 & +\left(\mathbb{I}-C_{{\rm mirror}}\right)^{n}\left(\mathbb{I}-C_{A_{L}\cap\bar{A}_{R}}\oplus C_{\bar{A}_{L}\cap A_{R}}\right)^{1-n}\bigg].\label{eq:PRMI-SPA-asymptotics}
\end{align}
}

\selectlanguage{american}%
Furthermore, due to the same logic presented in Appendix \ref{sec:Moments-for-two-intervals-appendix}
regarding similar moments of correlation matrices, ${\cal D}^{\left(n\right)}$
should be independent (at the leading order) of the distance between
\foreignlanguage{english}{$A_{{\scriptscriptstyle L}}\cap\bar{A}_{{\scriptscriptstyle R}}$
and $\bar{A}_{{\scriptscriptstyle L}}\cap A_{{\scriptscriptstyle R}}$,
meaning that we can regard this distance as infinite, as we have done
in Subsecs.~\ref{subsec:Asymptotics-of-entropies-derivation} and
\ref{subsec:Asymptotics-of-negativities-derivation}. Then, recalling
Eq.~(\ref{eq:Correlation-function-long-range}), we see that both
$C_{{\rm mirror}}$ and $C_{A_{L}\cap\bar{A}_{R}}\oplus C_{\bar{A}_{L}\cap A_{R}}$
are of a block-Toeplitz form: the former is represented by the block
symbol $\Phi$ given in Eq.~(\ref{eq:Block-Toeplitz-symbol-mirror}),
while the latter is represented by a block symbol $\Phi_{\times}$
given by $\left(\Phi_{\times}\right)_{ij}=\delta_{ij}\Phi_{ij}$.}

\selectlanguage{english}%
In general, a product of two block-Toeplitz matrices is not itself
a block-Toeplitz matrix. This fact prevents us from directly using
the Szeg\H{o}-Widom formula of Eq.~(\ref{eq:Szego-Widom formula})
to estimate the scaling of Eq.~(\ref{eq:PRMI-SPA-asymptotics}) (it
is also what led us to first express R\'enyi entropies and negativities
as in Eqs.~(\ref{eq:Entropy-from-determinants}) and (\ref{eq:Negativity-from-determinants}),
so that the Szeg\H{o}-Widom formula could be in fact employed). Instead,
we use a recent conjecture that generalizes the Szeg\H{o}-Widom formula.
Let $\Psi_{1},\Upsilon_{1},\ldots,\Psi_{p},\Upsilon_{p}$ be block
symbols of the same size, and let $T_{\ell}\!\left[\Psi_{j}\right],T_{\ell}\!\left[\Upsilon_{j}\right]$
($j=1,\ldots,p$) be the block-Toeplitz matrices generated by them
(each comprised of $\ell^{2}$ blocks). Then, it was conjectured in
Ref.~\citep{10.21468/SciPostPhys.15.3.089} that 
\begin{align}
 & \ln\det\!\left[\mathbb{I}+\prod_{j=1}^{p}T_{\ell}\!\left[\Psi_{j}\right]\left(T_{\ell}\!\left[\Upsilon_{j}\right]\right)^{-1}\right]\nonumber \\
 & \sim\ell\int_{-\pi}^{\pi}\frac{dk}{2\pi}\ln\det\!\left[\mathbb{I}+\prod_{j=1}^{p}\Psi_{j}{\Upsilon_{j}}^{-1}\right].
\end{align}

Applying this conjecture to Eq.~(\ref{eq:PRMI-SPA-asymptotics}),
we obtain
\begin{align}
{\cal D}^{\left(n\right)} & \sim\frac{\ell_{{\rm mirror}}}{n-1}\int_{-\pi}^{\pi}\frac{dk}{2\pi}\ln\det\!\bigg[\!\left(\Phi\right)^{n}\left(\Phi_{\times}\right)^{1-n}\nonumber \\
 & +\left(\mathbb{I}-\Phi\right)^{n}\left(\mathbb{I}-\Phi_{\times}\right)^{1-n}\bigg].\label{eq:PRMI-SW-asymptotics}
\end{align}
It is not difficult to express the integer powers of $\Phi\!\left(k\right)$
and $\Phi_{\times}\!\left(k\right)$: the latter is diagonal for all
$k$, and the former is diagonal for $k<0$, while for $k>0$ one
may check that
\begin{equation}
\left(\Phi\right)^{n}=\left(\begin{array}{cc}
{\cal R}{f_{{\scriptscriptstyle L}}}^{n}+{\cal T}{f_{{\scriptscriptstyle R}}}^{n} & t_{{\scriptscriptstyle L}}r_{{\scriptscriptstyle L}}^{*}\left({f_{{\scriptscriptstyle L}}}^{n}-{f_{{\scriptscriptstyle R}}}^{n}\right)\\
t_{{\scriptscriptstyle L}}^{*}r_{{\scriptscriptstyle L}}\left({f_{{\scriptscriptstyle L}}}^{n}-{f_{{\scriptscriptstyle R}}}^{n}\right) & {\cal T}{f_{{\scriptscriptstyle L}}}^{n}+{\cal R}{f_{{\scriptscriptstyle R}}}^{n}
\end{array}\right).
\end{equation}
Therefore, by substituting the explicit expressions for $\Phi$ and
$\Phi_{\times}$ into Eq.~(\ref{eq:PRMI-SW-asymptotics}), it is
straightforward to finally arrive at Eq.~(\ref{eq:PRMI-main-result}).

\section{Summary and outlook\label{sec:Summary-and-outlook}}

This paper advances our program for exploring the correlation and
entanglement structure of nonequilibrium steady states, focusing on
the steady state of biased free fermions in one dimension in the presence
of a homogeneity-breaking impurity. The results reported here extend
those of Ref.~\citep{fraenkel2022extensive}, by showing that the
phenomenon of stationary volume-law long-range entanglement arises
not only for a chemical-potential bias at zero temperature, but more
generally given any difference between the equilibrium distributions
of the two edge reservoirs that impose the nonequilibrium bias. In
particular, this strong long-range entanglement is found to be robust
even at finite temperatures. We offered an intuitive explanation for
the source of the phenomenon, attributing it to the coherence between
the transmitted and reflected parts of wavepackets occupying each
single-particle energy mode, provided that the two scattering states
that correspond to the same energy are not simultaneously occupied.

Our analysis produced exact leading-order asymptotic expressions for
various bipartite quantum information quantities, computed for two
subsystems located on opposite sides of the impurity. Most notably,
this includes exact formulas for the fermionic negativity (quantifying
entanglement) and the mutual information (quantifying the combined
classical and quantum correlations) between the two subsystems. Additional
analytical expressions were derived for R\'enyi negativities, the
R\'enyi mutual information, and the Petz-R\'enyi mutual information,
where the latter is itself a proper measure of inter-subsystem correlations.
All of these analytical results were obtained without assuming a specific
structure of the impurity (other than requiring that it will be quadratic
and charge-conserving), and are expressed solely in terms of the scattering
probabilities associated with the impurity, and of the equilibrium
energy distributions of the reservoirs.

The exact expressions for the volume-law terms of the negativity and
the mutual information vanish either in the absence of the nonequilibrium
bias, or if the impurity perfectly transmits or perfectly reflects
each particle, but are positive otherwise. In this sense, the model
we studied can be regarded as a minimal model giving rise to volume-law
long-range entanglement, seeing that the bias and the locally broken
homogeneity, which are essentially the defining features of the model,
constitute the necessary and sufficient conditions for the phenomenon.
The analytical expression for the mutual information also nicely captures
the local nonequilibrium energy distribution that arises due to the
combination of the bias and the scattering, as indeed required in
order to violate the thermal equilibrium area law of the mutual information.

Furthermore, we found a proportionality relation (at the leading volume-law
order) between the negativity, the $\frac{1}{2}$-RMI and the $\frac{5}{4}$-PRMI,
which holds at zero temperature but breaks down otherwise. As we elaborated
in Sec.~\ref{sec:Results}, this relation suggests that the strong
long-range correlations are entirely of a quantum coherent nature,
while its breakdown signals a significant contribution to these correlations
beyond what is quantified by the negativity. Like in other free-fermion
models where this relation was observed to break down \citep{alba2022logarithmic,Caceffo_2023,PhysRevB.107.075157},
this effect occurs in our case when the global state of the system
is mixed; this property is however insufficient in general, see e.g.~Ref.~\citep{PhysRevB.106.024304}.
The precise conditions for the breakdown of this proportionality relation,
as well as its operational meaning, remain to the best of our knowledge
subjects of open questions.

We stress that while, for concreteness, we treated a particular choice
for the energy dispersion relation, our analysis does not depend on
it, and should be applicable to any gapless free-fermion model. Similarly,
given that our results are expressed directly using the energy distribution
functions of the edge reservoirs, more general forms can be chosen
for these distributions, and they need not necessarily be of the Fermi-Dirac
form.

Along with the physical insights produced by this work, it also showcases
the versatility of the analytical machinery that we put forward in
Ref.~\citep{fraenkel2022extensive}, and used here to generalize
our original results. It is reasonable to expect that the same method
would be useful for similar investigations concerning different noninteracting
fermionic models. This method should also be appropriate for calculations
of quantities that address the interplay of charge conservation with
steady-state correlations across the impurity, namely the full counting
statistics \citep{doi:10.1063/1.3149497,10.21468/SciPostPhys.8.3.036,PhysRevLett.131.140401}
and charge-resolved entanglement measures \citep{PhysRevLett.120.200602,PhysRevA.98.032302,PhysRevB.100.235146,Bonsignori_2019,Fraenkel_2020,10.21468/SciPostPhys.10.5.111,Parez_2021,Zhao2021}.

Another worthwhile pursuit could be the analytical study of subleading
corrections beyond the volume-law asymptotics derived here. Indeed,
such corrections often have universal forms that disclose fundamental
physical attributes of the system in question \citep{Calabrese_2004,PhysRevLett.96.010404,PhysRevLett.96.110404}.
In Ref.~\citep{fraenkel2023exact} we already derived the exact forms
of the logarithmic corrections to the leading terms of correlation
measures in the case of zero temperature, finding that they can become
comparable to the volume-law terms when the chemical-potential bias
is small enough. While the method used in Ref.~\citep{fraenkel2023exact}
is not immediately suitable to tackle the case of finite temperatures,
we note that the zero-temperature exact expressions are similar in
their structure to the asymptotic expressions for the same quantities
in ground states of conformal field theories \citep{Calabrese_2009,Calabrese_disjoint_2009}.
Thus, the well-known finite-temperature scaling in the case of the
latter (which is logarithmic in the temperature) could presumably
provide guidance in calculating subleading contributions in the case
of the nonequilibrium steady state.

On a more ambitious note, we reiterate the resemblance between the
stationary behavior of the MI and the negativity studied here and
the transient long-range peak of these quantities following a nonequilibrium
quench of closed integrable systems \citep{Mesty_n_2017,10.21468/SciPostPhys.4.3.017,Alba_2019,PhysRevB.100.115150}.
Even though the latter include interacting models, this transient
phenomenon is explained by the quasiparticle picture, which portrays
information spreading as a process mediated by independent pairs of
counter-propagating quasiparticles \citep{doi:10.1073/pnas.1703516114,Calabrese_2005}.
It is tempting to conjecture that the long-range volume-law scaling
we reported here, considering an open biased system, would also apply
to integrable interacting systems containing an impurity \citep{rylands2023transport}.
The success or failure of this conjecture could offer a glimpse into
the limits of validity of the quasiparticle picture \citep{PhysRevX.12.031016,PhysRevB.106.L220304}.

The effects of integrability breaking, which in the case of the quench
scenario destroys the long-range coherence of information propagation
\citep{PhysRevB.100.115150}, are likewise an exciting prospective
subject for future research. Integrability could be broken either
by the impurity \citep{PhysRevB.98.235128,Bastianello_2019,PhysRevLett.125.070605}
or in the bulk of the model, and each case could potentially lead
to a quantitatively different scaling of quantum correlation measures.
We may mention nonequilibrium bosonization \citep{PhysRevB.81.085436,PhysRevLett.105.256802},
which was designed to treat biased one-dimensional systems of interacting
fermions, as a technique through which it might be possible to address
such problems.

The impact of decoherence and dissipation on the phenomenon that we
described certainly merits a separate consideration as well \citep{10.21468/SciPostPhys.12.1.011,PhysRevLett.129.056802,PhysRevA.85.012324},
as do possible similar studies applied to steady states of conformal
field theories \citep{10.21468/SciPostPhys.14.4.070,Erdmenger2017,10.21468/SciPostPhys.11.3.047}.
Finally, we note that these various questions can be studied experimentally
using any quantum simulation platform that offers local density resolution
\citep{Islam2015,doi:10.1126/science.aaf6725,Tajik2023}.
\begin{acknowledgments}
We thank Johanna Erdmenger and Ren\'e Meyer for discussions that
motivated this work. We also thank Filiberto Ares, Colin Rylands,
Pasquale Calabrese, Eran Sela and Yuval Gefen for helpful discussions.
Our work was supported by the Israel Science Foundation (ISF) and
the Directorate for Defense Research and Development (DDR\&D) grant
No.~3427/21, by the ISF grant No.~1113/23, and by the US-Israel
Binational Science Foundation (BSF) grant No.~2020072. S.F.~is grateful
for the support of the Azrieli Foundation Fellows program.
\end{acknowledgments}

\appendix

\section{Moments of correlation matrices restricted to $A_{{\scriptscriptstyle L}}$
and $A_{{\scriptscriptstyle R}}$\label{sec:Moments-for-single-intervals-appendix}}

In this appendix we detail the method used to compute integer moments
of $C_{A_{L}}$ and $C_{A_{R}}$ (the two-point correlation matrices
restricted to $A_{{\scriptscriptstyle L}}$ and $A_{{\scriptscriptstyle R}}$,
respectively), leading to Eq.~(\ref{eq:Correlation-matrix-moment-interval-asymptotics}).
The method relies on the stationary phase approximation (SPA) \citep{doi:10.1137/1.9780898719260},
and the calculation is almost identical to the one performed in the
zero-temperature case, which was detailed in Ref.~\citep{fraenkel2022extensive};
we provide the details of the calculation in the more general case
for the sake of completeness. We focus our discussion on moments of
$C_{A_{R}}$, with the corresponding computation for $C_{A_{L}}$
being virtually equivalent.

We begin by introducing the following notations:\begin{widetext}
\begin{align}
\Xi^{00}\!\left(k_{j-1},k_{j}\right) & =t_{{\scriptscriptstyle L}}\!\left(\left|k_{j-1}\right|\right)t_{{\scriptscriptstyle L}}^{*}\!\left(\left|k_{j}\right|\right){\cal W}_{{\scriptscriptstyle R}}\!\left(\frac{k_{j-1}-k_{j}}{2}\right)\int_{0}^{1}d\xi\,e{}^{i\ell_{R}\left(k_{j-1}-k_{j}\right)\xi},\nonumber \\
\Xi^{11}\!\left(k_{j-1},k_{j}\right) & =\int_{0}^{1}d\xi\left\{ {\cal W}_{{\scriptscriptstyle R}}\!\left(\frac{k_{j}-k_{j-1}}{2}\right)e{}^{i\ell_{R}\left(k_{j}-k_{j-1}\right)\xi}+r_{{\scriptscriptstyle R}}\!\left(\left|k_{j-1}\right|\right)r_{{\scriptscriptstyle R}}^{*}\!\left(\left|k_{j}\right|\right){\cal W}_{{\scriptscriptstyle R}}\!\left(\frac{k_{j-1}-k_{j}}{2}\right)e^{i\ell_{R}\left(k_{j-1}-k_{j}\right)\xi}\right\} \nonumber \\
 & +\int_{0}^{1}d\xi\left\{ r_{{\scriptscriptstyle R}}^{*}\!\left(\left|k_{j}\right|\right){\cal W}_{{\scriptscriptstyle R}}\!\left(\frac{-k_{j-1}-k_{j}}{2}\right)e{}^{-i\ell_{R}\left(k_{j-1}+k_{j}\right)\xi}+r_{{\scriptscriptstyle R}}\!\left(\left|k_{j-1}\right|\right){\cal W}_{{\scriptscriptstyle R}}\!\left(\frac{k_{j-1}+k_{j}}{2}\right)e^{i\ell_{R}\left(k_{j-1}+k_{j}\right)\xi}\right\} ,\nonumber \\
\Xi^{01}\!\left(k_{j-1},k_{j}\right) & =t_{{\scriptscriptstyle L}}\!\left(\left|k_{j-1}\right|\right)\int_{0}^{1}d\xi\left\{ {\cal W}_{{\scriptscriptstyle R}}\!\left(\frac{k_{j-1}+k_{j}}{2}\right)e{}^{i\ell_{R}\left(k_{j-1}+k_{j}\right)\xi}+r_{{\scriptscriptstyle R}}^{*}\!\left(\left|k_{j}\right|\right){\cal W}_{{\scriptscriptstyle R}}\!\left(\frac{k_{j-1}-k_{j}}{2}\right)e^{i\ell_{R}\left(k_{j-1}-k_{j}\right)\xi}\right\} ,\label{eq:SPA-symbols}
\end{align}
\end{widetext}where ${\cal W}_{{\scriptscriptstyle R}}\!\left(x\right)={\cal W}_{m_{0}+d_{R}}\!\left(x\right)$
(recall the definition of ${\cal W}_{r}$, given just below Eq.~(\ref{eq:Sum-as-integral-identity})).
We also define the functions
\begin{align}
\widetilde{\Xi}^{00}\!\left(k_{j-1},k_{j}\right) & ={\cal T}\!\left(\left|k_{j}\right|\right)\int_{0}^{1}\!d\xi\,e{}^{i\ell_{R}\left(k_{j-1}-k_{j}\right)\xi},\nonumber \\
\widetilde{\Xi}^{11}\!\left(k_{j-1},k_{j}\right) & =\int_{0}^{1}\!d\xi\bigg\{ e^{i\ell_{R}\left(k_{j}-k_{j-1}\right)\xi}\nonumber \\
 & +{\cal R}\!\left(\left|k_{j}\right|\right)e^{i\ell_{R}\left(k_{j-1}-k_{j}\right)\xi}\bigg\},\nonumber \\
\widetilde{\Xi}^{01}\!\left(k_{j-1},k_{j}\right) & =t_{{\scriptscriptstyle L}}\!\left(\left|k_{j}\right|\right)r_{{\scriptscriptstyle R}}^{*}\!\left(\left|k_{j}\right|\right)\int_{0}^{1}\!d\xi\,e^{i\ell_{R}\left(k_{j-1}-k_{j}\right)\xi}\,,\label{eq:SPA-symbols-leading-order}
\end{align}
as well as $\Xi^{10}\!\left(k_{j-1},k_{j}\right)=\Xi^{01}\!\left(k_{j},k_{j-1}\right)^{*}$
and $\widetilde{\Xi}^{10}\!\left(k_{j-1},k_{j}\right)=\widetilde{\Xi}^{01}\!\left(k_{j},k_{j-1}\right)^{*}$.
Next, we substitute the eigenstate wavefunctions in Eqs.~(\ref{eq:Left-scattering-states})
and (\ref{eq:Right-scattering-states}) into the general expression
for a correlation matrix moment in Eq.~(\ref{eq:Correlation-matrix-moment-general}).
Using the identity given in Eq.~(\ref{eq:Sum-as-integral-identity}),
this yields
\begin{align}
{\rm Tr}\!\left[\left(C_{A_{R}}\right)^{p}\right] & ={\ell_{{\scriptscriptstyle R}}}^{p}\!\!\!\!\!\!\int_{\left[-\pi,\pi\right]^{p}}\frac{d^{p}k}{\left(2\pi\right)^{p}}\prod_{j=1}^{p}\widetilde{f}\!\left(k_{j}\right)\nonumber \\
 & \times\!\!\!\sum_{\overrightarrow{a}\in\left\{ 0,1\right\} ^{\otimes p}}\prod_{j=1}^{p}\left[\Xi^{a_{j-1}a_{j}}\!\left(k_{a_{j-1}},k_{a_{j}}\right)\Theta\!\left(k_{a_{j}}\right)\right],\label{eq:Correlation-matrix-moment-SPA-exact}
\end{align}
where $\Theta\!\left(x\right)$ is the Heaviside step function, and
we defined $k_{a_{j}}=\left(-1\right)^{a_{j}}k_{j}$. It is readily
seen that Eq.~(\ref{eq:Correlation-matrix-moment-SPA-exact}) can
be cast in the general form given in Eq.~(\ref{eq:Correlation-matrix-moment-SPA-form}),
and therefore the result of Eq.~(\ref{eq:Correlation-matrix-moment-SPA-asymptotics}),
which stems from the SPA, can be directly applied to obtain the leading-order
asymptotics of the correlation matrix moment.

Eq\@.~(\ref{eq:Correlation-matrix-moment-SPA-asymptotics}) dictates
that the summands ${\cal G}\!\left(\overrightarrow{\tau},\overrightarrow{\sigma}\right)$
in Eq.~(\ref{eq:Correlation-matrix-moment-SPA-form}) that contribute
to the leading order are only those with $\overrightarrow{\tau}=\overrightarrow{\sigma}$.
Furthermore, it constrains the integration variable $\overrightarrow{k}$
in the integral expression for ${\cal G}\!\left(\overrightarrow{\sigma},\overrightarrow{\sigma}\right)$
(see Eq.~(\ref{eq:Correlation-matrix-moment-SPA-form-summands}))
to a subdomain where $k_{\sigma_{j-1}}=k_{\sigma_{j}}$ for $1\le j\le p$.
This constraint entails a considerable simplification of Eq.~(\ref{eq:Correlation-matrix-moment-SPA-exact})
when considering it only up to the leading order. Indeed, the aforementioned
integration subdomain would eventually correspond to the vanishing
of the terms $k_{j-1}\pm k_{j}$ appearing inside the exponents in
Eq.~(\ref{eq:SPA-symbols}). However, the term $k_{j-1}+k_{j}$ does
not vanish if $k_{j-1},k_{j}>0$, while each term $\Xi^{a_{j-1}a_{j}}\!\left(k_{a_{j-1}},k_{a_{j}}\right)\Theta\!\left(k_{a_{j-1}}\right)\Theta\!\left(k_{a_{j}}\right)$
in Eq.~(\ref{eq:Correlation-matrix-moment-SPA-exact}) vanishes due
to the step functions unless $k_{a_{j-1}},k_{a_{j}}>0$. We may therefore
omit from the functions defined in Eq.~(\ref{eq:SPA-symbols}) the
integrals where $k_{j-1}+k_{j}$ appears in the exponent. Due to the
same subdomain constraint, we may also substitute $k_{j-1}=k_{j}$
into all $\xi$-independent factors appearing in Eq.~(\ref{eq:SPA-symbols}).
This step is captured by replacing in Eq.~(\ref{eq:Correlation-matrix-moment-SPA-exact})
the functions $\Xi^{a_{j-1}a_{j}}$ with the functions $\widetilde{\Xi}^{a_{j-1}a_{j}}$,
where the latter were defined in Eq.~(\ref{eq:SPA-symbols-leading-order})
(note that we used the fact that ${\cal W}_{r}\!\left(0\right)=1$).

Therefore, to a leading order we may write
\begin{align}
{\rm Tr}\!\left[\left(C_{A_{R}}\right)^{p}\right] & \sim{\ell_{{\scriptscriptstyle R}}}^{p}\!\!\!\!\!\!\int_{\left[-\pi,\pi\right]^{p}}\frac{d^{p}k}{\left(2\pi\right)^{p}}\prod_{j=1}^{p}\widetilde{f}\!\left(k_{j}\right)\nonumber \\
 & \times\!\!\!\sum_{\overrightarrow{a}\in\left\{ 0,1\right\} ^{\otimes p}}\prod_{j=1}^{p}\left[\widetilde{\Xi}^{a_{j-1}a_{j}}\!\left(k_{a_{j-1}},k_{a_{j}}\right)\Theta\!\left(k_{a_{j}}\right)\right].
\end{align}
Casting this in the form of Eq.~(\ref{eq:Correlation-matrix-moment-SPA-form})
and omitting integrals ${\cal G}\!\left(\overrightarrow{\tau},\overrightarrow{\sigma}\right)$
with $\overrightarrow{\tau}\neq\overrightarrow{\sigma}$, we have\begin{widetext}
\begin{align}
{\rm Tr}\!\left[\left(C_{A_{R}}\right)^{p}\right] & \sim{\ell_{{\scriptscriptstyle R}}}^{p}\!\!\!\int_{\left[-\pi,0\right]^{p}}\!\!\!\frac{d^{p}k}{\left(2\pi\right)^{p}}\!\!\int_{\left[0,1\right]^{p}}\!\!\!d^{p}\xi\prod_{j=1}^{p}\!\widetilde{f}\!\left(k_{j}\right)e^{i\ell_{R}\left(k_{j-1}-k_{j}\right)\xi_{j}}+{\ell_{{\scriptscriptstyle R}}}^{p}\!\!\!\int_{\left[-\pi,\pi\right]^{p}}\!\!\!\frac{d^{p}k}{\left(2\pi\right)^{p}}\prod_{j=1}^{p}\!\widetilde{f}\!\left(k_{j}\right)\nonumber \\
 & \times\int_{\left[0,1\right]^{p}}\!\!\!d^{p}\xi\sum_{\overrightarrow{a}\in\left\{ 0,1\right\} ^{\otimes p}}\prod_{j=1}^{p}\left\{ \Theta\!\left(k_{a_{j}}\right)\exp\!\left[i\ell_{{\scriptscriptstyle R}}\!\left(k_{a_{j-1}}\!-\!k_{a_{j}}\right)\!\xi_{j}\right]\frac{1+\left(-1\right)^{a_{j}}\!\left[{\cal T}\!\left(k_{a_{j}}\right)\!-\!{\cal R}\!\left(k_{a_{j}}\right)\right]}{2}\right\} ,
\end{align}
\end{widetext}which then, using Eq.~(\ref{eq:Correlation-matrix-moment-SPA-asymptotics}),
leads to the result
\begin{align}
{\rm Tr}\!\left[\left(C_{A_{R}}\right)^{p}\right] & \sim\ell_{{\scriptscriptstyle R}}\int_{-\pi}^{0}\frac{dk}{2\pi}\left(\widetilde{f}\!\left(k\right)\right)^{p}\nonumber \\
 & +\ell_{{\scriptscriptstyle R}}\int_{0}^{\pi}\frac{dk}{2\pi}\left({\cal T}\!\left(k\right)\widetilde{f}\!\left(k\right)+{\cal R}\!\left(k\right)\widetilde{f}\!\left(-k\right)\right)^{p}.\label{eq:Correlation-matrix-moment-SPA-final}
\end{align}
As mentioned before, the calculation of the moments of $C_{A_{L}}$
is equivalent, yielding a result similar to Eq.~(\ref{eq:Correlation-matrix-moment-SPA-final}),
up to replacing $\ell_{{\scriptscriptstyle R}}$ with $\ell_{{\scriptscriptstyle L}}$
and $\widetilde{f}\!\left(k\right)$ with $\widetilde{f}\!\left(-k\right)$.
This can also be written in the form given in Eq.~(\ref{eq:Correlation-matrix-moment-interval-asymptotics}).

\section{Stationary phase approximation for moments of correlation matrices
restricted to $A$\label{sec:Moments-for-two-intervals-appendix}}

In this appendix, we lay out the argument that leads to the decomposition
of correlation matrix moments given in Eqs.~(\ref{eq:Correlation-matrix-moment-combined-decomposition})
and (\ref{eq:Joint-moment-SPA-decomposition}) for the subsystem $A=A_{{\scriptscriptstyle L}}\cup A_{{\scriptscriptstyle R}}$.
For this purpose, we assume that the length scales $\Delta\ell_{{\scriptscriptstyle L}}$,
$\Delta\ell_{{\scriptscriptstyle R}}$ and $\ell_{{\rm mirror}}$
are all large, and that they scale linearly with the same large parameter
$\ell$. We define $\Delta\ell_{-}=\left|d_{{\scriptscriptstyle L}}-d_{{\scriptscriptstyle R}}\right|$
and $\Delta\ell_{+}=\left|\ell_{{\scriptscriptstyle L}}+d_{{\scriptscriptstyle L}}-\ell_{{\scriptscriptstyle R}}-d_{{\scriptscriptstyle R}}\right|$;
note that $\Delta\ell_{\pm}$ correspond to $\Delta\ell_{{\scriptscriptstyle L}}$
and $\Delta\ell_{{\scriptscriptstyle R}}$, with the specific correspondence
depending on the relative positions of $\bar{A}_{{\scriptscriptstyle L}}$
(the mirror image of $A_{{\scriptscriptstyle L}}$) and $A_{{\scriptscriptstyle R}}$.
We also write $\ell_{{\rm mirror}}=\alpha_{{\rm m}}\ell$ and $\Delta\ell_{\pm}=\alpha_{\pm}\ell$.

The core of the argument is related to the large-$\ell$ leading-order
asymptotics of integrals of the following form (adopting the notation
$k_{a_{j}}=\left(-1\right)^{a_{j}}k_{j}$ from Appendix \ref{sec:Moments-for-single-intervals-appendix}):
\begin{align}
{\cal G}_{\overrightarrow{\alpha}}\!\left(\overrightarrow{\tau},\overrightarrow{\sigma}\right) & =\left[\prod_{j=1}^{p}\left(\alpha_{j}\ell\right)\right]\int_{\left[-\pi,\pi\right]^{p}}\frac{d^{p}k}{\left(2\pi\right)^{p}}\int_{\left[0,1\right]^{p}}d^{p}\xi\,g\!\left(\overrightarrow{k}\right)\nonumber \\
 & \times\exp\!\left[i\ell\sum_{j=1}^{p}\left(k_{\tau_{j-1}}-k_{\sigma_{j}}\right)\left(\alpha_{j}\xi_{j}+\beta_{j}\right)\right].\label{eq:SPA-summand-combined-subsystem}
\end{align}
Here, $\left(\alpha_{j},\beta_{j}\right)\in\left\{ \left(\alpha_{-},0\right),\left(\alpha_{{\rm m}},\alpha_{-}\right),\left(\alpha_{+},\alpha_{-}+\alpha_{{\rm m}}\right)\right\} $
for each $1\le j\le p$, and the form of the function $g$ might depend
on $\overrightarrow{\tau}$, $\overrightarrow{\sigma}$ or $\overrightarrow{\alpha}$,
but not on $\ell$. In Ref.~\citep{fraenkel2022extensive} we have
found, based on the SPA, that an integral of the form of ${\cal G}_{\overrightarrow{\alpha}}\!\left(\overrightarrow{\tau},\overrightarrow{\sigma}\right)$
can scale linearly with $\ell$ only if $\alpha_{1}=\alpha_{2}=\ldots=\alpha_{p}$;
otherwise, $\lim_{\ell\to\infty}{\cal G}_{\overrightarrow{\alpha}}/\ell=0$,
even when $\overrightarrow{\tau}=\overrightarrow{\sigma}$. Intuitively,
this happens because when $\overrightarrow{\tau}=\overrightarrow{\sigma}$
and all the $\alpha_{j}$ are the same, then the stationary points
of the function inside the exponent in Eq.~(\ref{eq:SPA-summand-combined-subsystem})
constitute the entire line $\xi_{1}=\xi_{2}=\ldots=\xi_{p}$, while
they are reduced, at most, to an isolated point in $\overrightarrow{\xi}$-space
if $\alpha_{j-1}\neq\alpha_{j}$.

Consider now the integer moments of the matrix $C_{A}$, expressed
as in Eq.~(\ref{eq:Correlation-matrix-moment-general}). The sum
over $m\in A$ in Eq.~(\ref{eq:Correlation-matrix-moment-general})
can be split in the following way:
\begin{align}
 & \sum_{m\in A}\left\langle m|k_{j-1}\right\rangle \!\left\langle k_{j}|m\right\rangle \nonumber \\
 & =\sum_{m\in\bar{A}_{L}\cap A_{R}}\!\!\!\!\!\!\left[\left\langle -m|k_{j-1}\right\rangle \!\left\langle k_{j}|\!-\!m\right\rangle +\left\langle m|k_{j-1}\right\rangle \!\left\langle k_{j}|m\right\rangle \right]\nonumber \\
 & +\sum_{m\in\bar{A}_{L}\setminus A_{R}}\!\!\!\!\!\!\left\langle -m|k_{j-1}\right\rangle \!\left\langle k_{j}|\!-\!m\right\rangle +\!\!\!\!\sum_{m\in A_{R}\setminus\bar{A}_{L}}\!\!\!\!\!\!\left\langle m|k_{j-1}\right\rangle \!\left\langle k_{j}|m\right\rangle .\label{eq:Site-sum-decomposition-A}
\end{align}
Next, we define $d_{-}=\min\!\left\{ d_{{\scriptscriptstyle L}},d_{{\scriptscriptstyle R}}\right\} $,
$d_{+}=\max\!\left\{ d_{{\scriptscriptstyle L}},d_{{\scriptscriptstyle R}}\right\} $
and ${\cal W}_{-}\!\left(x\right)={\cal W}_{m_{0}+d_{-}}\!\left(x\right)$
(recall the definition of ${\cal W}_{r}$, given just below Eq.~(\ref{eq:Sum-as-integral-identity})).
After substituting the eigenstate wavefunctions of Eqs.~(\ref{eq:Left-scattering-states})--(\ref{eq:Right-scattering-states})
into the first sum on the right-hand side of Eq.~(\ref{eq:Site-sum-decomposition-A}),
this sum can be brought to an integral form using the identity
\begin{align}
\sum_{m=m_{0}+d_{+}+1}^{m_{0}+d_{+}+\ell_{{\rm mirror}}}e^{im\left(k_{j-1}-k_{j}\right)} & =\alpha_{{\rm m}}\ell\,{\cal W}_{-}\!\!\left(\frac{k_{j-1}-k_{j}}{2}\right)\nonumber \\
 & \times\int_{0}^{1}d\xi\,e^{i\ell\left(k_{j-1}-k_{j}\right)\left(\alpha_{{\rm m}}\xi+\alpha_{-}\right)},\label{eq:Sum-as-integral-mirroring-sites}
\end{align}
which is equivalent to Eq.~(\ref{eq:Sum-as-integral-identity}).
Similarly, for the two other sums appearing on the right-hand side
of Eq.~(\ref{eq:Site-sum-decomposition-A}), we may use Eq.~(\ref{eq:Sum-as-integral-identity})
directly with $r=m_{0}+d_{-}$ and $s=\Delta\ell_{-}$, as well as
the fact that
\begin{align}
 & \sum_{m=m_{0}+d_{+}+\ell_{{\rm mirror}}+1}^{m_{0}+d_{+}+\ell_{{\rm mirror}}+\Delta\ell_{+}}\!\!\!\!e^{im\left(k_{j-1}-k_{j}\right)}\nonumber \\
 & =\alpha_{+}\ell\,{\cal W}_{-}\!\!\left(\frac{k_{j-1}-k_{j}}{2}\right)\int_{0}^{1}d\xi\,e^{i\ell\left(k_{j-1}-k_{j}\right)\left(\alpha_{+}\xi+\alpha_{-}+\alpha_{{\rm m}}\right)}.
\end{align}

In all, it is clear that the substitution of the wavefunctions from
Eqs.~(\ref{eq:Left-scattering-states})--(\ref{eq:Right-scattering-states})
into the expression in Eq.~(\ref{eq:Correlation-matrix-moment-general})
leads to the $p$th moment of $C_{A}$ being written as a sum of integrals
of the form given in Eq.~(\ref{eq:SPA-summand-combined-subsystem}).
The observation we cited from Ref.~\citep{fraenkel2022extensive}
means that we can ignore all of the integrals where the entries of
$\overrightarrow{\alpha}$ are not all the same, such that to a leading
order we have\begin{widetext}
\begin{align}
{\rm Tr}\!\left[\left(C_{A}\right)^{p}\right] & \sim\int_{\left[-\pi,\pi\right]^{p}}\!\!\frac{d^{p}k}{\left(2\pi\right)^{p}}\prod_{j=1}^{p}\!\widetilde{f}\!\left(k_{j}\right)\!\left[\sum_{m\in\bar{A}_{L}\cap A_{R}}\!\!\!\!\!\!\left[\left\langle -m|k_{j-1}\right\rangle \!\left\langle k_{j}|\!-\!m\right\rangle +\left\langle m|k_{j-1}\right\rangle \!\left\langle k_{j}|m\right\rangle \right]\right]\nonumber \\
 & +\int_{\left[-\pi,\pi\right]^{p}}\!\!\frac{d^{p}k}{\left(2\pi\right)^{p}}\prod_{j=1}^{p}\!\widetilde{f}\!\left(k_{j}\right)\!\left[\sum_{m\in\bar{A}_{L}\setminus A_{R}}\!\!\!\!\!\!\left\langle -m|k_{j-1}\right\rangle \!\left\langle k_{j}|\!-\!m\right\rangle \right]+\int_{\left[-\pi,\pi\right]^{p}}\!\!\frac{d^{p}k}{\left(2\pi\right)^{p}}\prod_{j=1}^{p}\!\widetilde{f}\!\left(k_{j}\right)\!\left[\sum_{m\in A_{R}\setminus\bar{A}_{L}}\!\!\!\!\!\!\left\langle m|k_{j-1}\right\rangle \!\left\langle k_{j}|m\right\rangle \right],\label{eq:Correlation-matrix-moment-decomposition-integral}
\end{align}
or simply
\begin{align}
{\rm Tr}\!\left[\left(C_{A}\right)^{p}\right] & \sim{\rm Tr}\!\left[\left(C_{{\rm mirror}}\right)^{p}\right]+{\rm Tr}\!\left[\left(C_{A_{L}\setminus\bar{A}_{R}}\right)^{p}\right]+{\rm Tr}\!\left[\left(C_{A_{R}\setminus\bar{A}_{L}}\right)^{p}\right].
\end{align}
Since $A_{{\scriptscriptstyle L}}\setminus\bar{A}_{{\scriptscriptstyle R}}$
and $A_{{\scriptscriptstyle R}}\setminus\bar{A}_{{\scriptscriptstyle L}}$
are just portions of $A_{{\scriptscriptstyle L}}$ and $A_{{\scriptscriptstyle R}}$
with lengths $\Delta\ell_{{\scriptscriptstyle L}}$ and $\Delta\ell_{{\scriptscriptstyle R}}$,
respectively, and given that the correlation matrix moments of these
subsystems scale linearly with their lengths, we have in fact arrived
at Eq.~(\ref{eq:Correlation-matrix-moment-combined-decomposition}).

Moreover, we may observe that the moment of $C_{{\rm mirror}}$ does
not depend (to the leading order) on the absolute distances of the
subsystems from the impurity. Indeed, if we use the identity in Eq.~(\ref{eq:Sum-as-integral-mirroring-sites})
when substituting the wavefunctions into the first integral in Eq.~(\ref{eq:Correlation-matrix-moment-decomposition-integral}),
we see that the dependence on the distance enters only through the
function ${\cal W}_{-}$. However, as we explained in Appendix \ref{sec:Moments-for-single-intervals-appendix},
the SPA always constrains the argument of that function to vanish
in its contribution to the leading-order term of the asymptotics.
Since ${\cal W}_{-}\!\left(0\right)=1$, we conclude that the leading-order
term does not depend on the distance.

Finally, we note that the argument that led to Eq.~(\ref{eq:Correlation-matrix-moment-decomposition-integral})
can be applied also to the joint moments given in Eq.~(\ref{eq:Joint-correlation-matrix-moment-general}).
Indeed, it is straightforward to check that by repeating the steps
that were taken in the case of the moments of $C_{A}$, one obtains
that, to a leading order,
\begin{align}
{\rm Tr}\!\left[C_{\gamma_{1}}\ldots C_{\gamma_{p}}\right] & \sim\int_{\left[-\pi,\pi\right]^{p}}\!\!\frac{d^{p}k}{\left(2\pi\right)^{p}}\prod_{j=1}^{p}\!\widetilde{f}\!\left(k_{j}\right)\!\left[\sum_{m\in\bar{A}_{L}\cap A_{R}}\!\!\!\!\!\!\left[\left(1-e^{\frac{2\pi i\gamma_{j}}{n}}\right)\left\langle -m|k_{j-1}\right\rangle \!\left\langle k_{j}|\!-\!m\right\rangle +\left(1+e^{\frac{-2\pi i\gamma_{j}}{n}}\right)\left\langle m|k_{j-1}\right\rangle \!\left\langle k_{j}|m\right\rangle \right]\right]\nonumber \\
 & +\int_{\left[-\pi,\pi\right]^{p}}\!\!\frac{d^{p}k}{\left(2\pi\right)^{p}}\prod_{j=1}^{p}\!\widetilde{f}\!\left(k_{j}\right)\!\left(1-e^{\frac{2\pi i\gamma_{j}}{n}}\right)\!\left[\sum_{m\in\bar{A}_{L}\setminus A_{R}}\!\!\!\!\!\!\left\langle -m|k_{j-1}\right\rangle \!\left\langle k_{j}|\!-\!m\right\rangle \right]\nonumber \\
 & +\int_{\left[-\pi,\pi\right]^{p}}\!\!\frac{d^{p}k}{\left(2\pi\right)^{p}}\prod_{j=1}^{p}\!\widetilde{f}\!\left(k_{j}\right)\!\left(1+e^{\frac{-2\pi i\gamma_{j}}{n}}\right)\!\left[\sum_{m\in A_{R}\setminus\bar{A}_{L}}\!\!\!\!\!\!\left\langle m|k_{j-1}\right\rangle \!\left\langle k_{j}|m\right\rangle \right].\label{eq:Joint-moment-decomposition-integral}
\end{align}
This yields Eq.~(\ref{eq:Joint-moment-SPA-decomposition}), again
by relying on the linear scaling with $\ell_{i}$ of moments of $C_{A_{i}}$.
As in the case of the moments of $C_{{\rm mirror}}$, the integral
in the first row of Eq.~(\ref{eq:Joint-moment-decomposition-integral})
is independent of the distance of the mirroring sites from the impurity,
by virtue of the same SPA argument.

\section{R\'enyi negativity from the Szeg\H{o}-Widom formula\label{sec:Renyi-negativity-appendix}}

Here we delineate the derivation of Eq.~(\ref{eq:Mirror-Renyi-negativity-asymptotics}),
which states the asymptotic scaling of the R\'enyi negativity between
two mirroring intervals of length $\ell_{{\rm mirror}}$, in the limit
of infinite distance between them and the impurity.

We begin by using \foreignlanguage{american}{the block-Toeplitz structure
of the matrix $\mathbb{I}-C_{{\rm mirror},\gamma}$, which is generated
by the block symbol $\mathbb{I}_{2}-\Phi_{\gamma}$, where $\Phi_{\gamma}$
is defined in Eq.~(\ref{eq:Block-Toeplitz-symbol-negativity}). The
Szeg\H{o}-Widom formula given in Eq.~(\ref{eq:Szego-Widom formula})
yields the asymptotics of the determinant of this matrix:}

\selectlanguage{american}%
\begin{align}
\frac{\ln\det\!\left[\mathbb{I}-C_{{\rm mirror},\gamma}\right]}{\ell_{{\rm mirror}}} & \sim\int_{0}^{\pi}\frac{dk}{2\pi}\bigg\{\ln\!\left[1\!-\!\left(1\!-\!e^{\frac{2\pi i\gamma}{n}}\right)\!f_{{\scriptscriptstyle L}}\right]+\ln\!\left[1\!-\!\left(1\!+\!e^{\frac{-2\pi i\gamma}{n}}\right)\!f_{{\scriptscriptstyle R}}\right]+\ln\!\bigg[1-\left(1\!-\!e^{\frac{2\pi i\gamma}{n}}\right)\left({\cal R}f_{{\scriptscriptstyle L}}\!+\!{\cal T}f_{{\scriptscriptstyle R}}\right)\nonumber \\
 & -\left(1\!+\!e^{\frac{-2\pi i\gamma}{n}}\right)\left({\cal T}f_{{\scriptscriptstyle L}}\!+\!{\cal R}f_{{\scriptscriptstyle R}}\right)+\left(1\!-\!e^{\frac{2\pi i\gamma}{n}}\right)\left(1\!+\!e^{\frac{-2\pi i\gamma}{n}}\right)f_{{\scriptscriptstyle L}}f_{{\scriptscriptstyle R}}\bigg]\bigg\}.
\end{align}

\selectlanguage{english}%
\end{widetext}\foreignlanguage{american}{Now, using the expression
in Eq.~(\ref{eq:Negativity-from-determinants}) for the R\'enyi
negativity in terms of determinants, we obtain
\begin{align}
{\cal E}_{{\rm mirror}}^{\left(n\right)} & \sim\ell_{{\rm mirror}}\int_{0}^{\pi}\!\!\frac{dk}{2\pi}\bigg\{\ln\!\left[\left(f_{{\scriptscriptstyle L}}\right)^{n}+\left(1-f_{{\scriptscriptstyle L}}\right)^{n}\right]\nonumber \\
 & +\ln\!\left[\left(f_{{\scriptscriptstyle R}}\right)^{n}+\left(1-f_{{\scriptscriptstyle R}}\right)^{n}\right]+\ln{\cal X}_{n}\!\left(k\right)\bigg\},\label{eq:Mirror-Renyi-negativity-from-SW}
\end{align}
where we introduced the notation
\begin{align}
{\cal X}_{n} & =\prod_{\gamma=-\frac{n-1}{2}}^{\frac{n-1}{2}}\bigg[1-\left(1-e^{\frac{2\pi i\gamma}{n}}\right)\left({\cal R}f_{{\scriptscriptstyle L}}+{\cal T}f_{{\scriptscriptstyle R}}\right)\nonumber \\
 & -\left(1+e^{\frac{-2\pi i\gamma}{n}}\right)\left({\cal T}f_{{\scriptscriptstyle L}}+{\cal R}f_{{\scriptscriptstyle R}}\right)\nonumber \\
 & +\left(1-e^{\frac{2\pi i\gamma}{n}}\right)\left(1+e^{\frac{-2\pi i\gamma}{n}}\right)f_{{\scriptscriptstyle L}}f_{{\scriptscriptstyle R}}\bigg].\label{eq:Negativity-auxiliary-polynomial-SW}
\end{align}
}

\selectlanguage{american}%
Comparing Eq.~(\ref{eq:Mirror-Renyi-negativity-from-SW}) to Eq.~(\ref{eq:Mirror-Renyi-negativity-asymptotics}),
we observe that what remains to be done is to prove the equality ${\cal X}_{n}\!\left(k\right)={\cal Y}_{n}\!\left(k\right)$,
where the definition of the function ${\cal Y}_{n}$ appears in Eq.~(\ref{eq:Negativity-auxiliary-polynomial}).
To do so, we will temporarily stop viewing $f_{{\scriptscriptstyle L}}$,
$f_{{\scriptscriptstyle R}}$ and ${\cal T}$ as functions of $k$,
and instead regard ${\cal X}_{n}$ and ${\cal Y}_{n}$ as polynomials
of degree $n$ in the variable ${\cal T}$ (recall that ${\cal R}=1-{\cal T}$),
with $f_{{\scriptscriptstyle L}}$ and $f_{{\scriptscriptstyle R}}$
being some fixed parameters. Recall that $n$ is taken to be an even
integer, such that ${\cal Y}_{n}$ is indeed a polynomial in ${\cal T}$
even though its definition formally contains non-integer powers of
${\cal T}$, as these powers cancel out when summing the different
terms in Eq.~(\ref{eq:Negativity-auxiliary-polynomial}).

To prove that ${\cal X}_{n}$ and ${\cal Y}_{n}$ are identical as
polynomials in ${\cal T}$, it suffices to show that they agree at
a certain point, say ${\cal T}=0$, and that they have the same $n$
roots when ${\cal T}$ is seen as a complex variable. The former requirement
is simpler: it is easy to check that substituting ${\cal T}=0$ into
both polynomials (i.e., into Eqs.~(\ref{eq:Negativity-auxiliary-polynomial})
and (\ref{eq:Negativity-auxiliary-polynomial-SW})) yields
\begin{align}
{\cal X}_{n}\!\left({\cal T}=0\right) & =\left[\left(f_{{\scriptscriptstyle L}}\right)^{n}+\left(1-f_{{\scriptscriptstyle L}}\right)^{n}\right]\left[\left(f_{{\scriptscriptstyle R}}\right)^{n}+\left(1-f_{{\scriptscriptstyle R}}\right)^{n}\right]\nonumber \\
 & ={\cal Y}_{n}\!\left({\cal T}=0\right).
\end{align}

As for the latter requirement regarding the equality between the roots
of the two polynomials, we may extract these roots using the fact
that in Eq.~(\ref{eq:Negativity-auxiliary-polynomial-SW}) ${\cal X}_{n}$
is already written as a product of terms that are linear in ${\cal T}$.
Its $n$ roots are thus given by
\begin{equation}
{\cal T}_{\gamma}=\frac{\left[1-\left(1-e^{\frac{2\pi i\gamma}{n}}\right)f_{{\scriptscriptstyle L}}\right]\left[1-\left(1+e^{\frac{-2\pi i\gamma}{n}}\right)f_{{\scriptscriptstyle R}}\right]}{\left(e^{\frac{2\pi i\gamma}{n}}+e^{\frac{-2\pi i\gamma}{n}}\right)\left(f_{{\scriptscriptstyle L}}-f_{{\scriptscriptstyle R}}\right)}.
\end{equation}
For convenience, we also define ${\cal R}_{\gamma}=1-{\cal T}_{\gamma}$.
Now we must show that ${\cal Y}_{n}\!\left({\cal T}_{\gamma}\right)=0$
for all $\gamma$. Indeed, one may check that the substitution of
${\cal T}_{\gamma}$ into the first two square-bracketed summands
in Eq.~(\ref{eq:Negativity-auxiliary-polynomial}) yields
\begin{align}
 & \left[{\cal T}_{\gamma}f_{{\scriptscriptstyle L}}+{\cal R}_{\gamma}f_{{\scriptscriptstyle R}}-f_{{\scriptscriptstyle L}}f_{{\scriptscriptstyle R}}\right]^{n}+\left[{\cal R}_{\gamma}f_{{\scriptscriptstyle L}}+{\cal T}_{\gamma}f_{{\scriptscriptstyle R}}-f_{{\scriptscriptstyle L}}f_{{\scriptscriptstyle R}}\right]^{n}\nonumber \\
 & =\left[\frac{1-\left(1-e^{\frac{2\pi i\gamma}{n}}\right)\left(f_{{\scriptscriptstyle L}}+f_{{\scriptscriptstyle R}}\right)-2e^{\frac{2\pi i\gamma}{n}}f_{{\scriptscriptstyle L}}f_{{\scriptscriptstyle R}}}{e^{\frac{2\pi i\gamma}{n}}+e^{\frac{-2\pi i\gamma}{n}}}\right]^{n}\nonumber \\
 & +\left[\frac{1-\left(1+e^{\frac{-2\pi i\gamma}{n}}\right)\left(f_{{\scriptscriptstyle L}}+f_{{\scriptscriptstyle R}}\right)+2e^{\frac{-2\pi i\gamma}{n}}f_{{\scriptscriptstyle L}}f_{{\scriptscriptstyle R}}}{e^{\frac{2\pi i\gamma}{n}}+e^{\frac{-2\pi i\gamma}{n}}}\right]^{n}.\label{eq:Appendix-polynomial-identity-1}
\end{align}
On the other hand, we observe that
\begin{align}
 & \left[\frac{1-f_{{\scriptscriptstyle L}}-f_{{\scriptscriptstyle R}}+2f_{{\scriptscriptstyle L}}f_{{\scriptscriptstyle R}}}{2}\right]^{2}+{\cal T}_{\gamma}{\cal R}_{\gamma}\left(f_{{\scriptscriptstyle L}}-f_{{\scriptscriptstyle R}}\right)^{2}\nonumber \\
 & =\left[\frac{\left(e^{\frac{-2\pi i\gamma}{n}}-e^{\frac{2\pi i\gamma}{n}}\right)\frac{1-f_{{\scriptscriptstyle L}}-f_{{\scriptscriptstyle R}}}{2}+f_{{\scriptscriptstyle L}}+f_{{\scriptscriptstyle R}}-2f_{{\scriptscriptstyle L}}f_{{\scriptscriptstyle R}}}{e^{\frac{2\pi i\gamma}{n}}+e^{\frac{-2\pi i\gamma}{n}}}\right]^{2},
\end{align}
which is equivalent to the statement that
\begin{align}
 & \sqrt{\left[\frac{1-f_{{\scriptscriptstyle L}}-f_{{\scriptscriptstyle R}}+2f_{{\scriptscriptstyle L}}f_{{\scriptscriptstyle R}}}{2}\right]^{2}+{\cal T}_{\gamma}{\cal R}_{\gamma}\left(f_{{\scriptscriptstyle L}}-f_{{\scriptscriptstyle R}}\right)^{2}}\nonumber \\
 & =\pm\frac{e^{\mp\frac{2\pi i\gamma}{n}}}{e^{\frac{2\pi i\gamma}{n}}+e^{\frac{-2\pi i\gamma}{n}}}\bigg[1-\left(1\mp e^{\pm\frac{2\pi i\gamma}{n}}\right)\left(f_{{\scriptscriptstyle L}}+f_{{\scriptscriptstyle R}}\right)\nonumber \\
 & \mp2e^{\pm\frac{2\pi i\gamma}{n}}f_{{\scriptscriptstyle L}}f_{{\scriptscriptstyle R}}\bigg]\mp\frac{1-f_{{\scriptscriptstyle L}}-f_{{\scriptscriptstyle R}}}{2}.\label{eq:Appendix-polynomial-identity-2}
\end{align}
Thus, using the identities in Eqs.~(\ref{eq:Appendix-polynomial-identity-1})
and (\ref{eq:Appendix-polynomial-identity-2}) as well as the fact
that $\left(e^{\frac{2\pi i\gamma}{n}}\right)^{n}=-1$ for all $\gamma$,
we conclude that in fact ${\cal Y}_{n}\!\left({\cal T}_{\gamma}\right)=0$
for all $\gamma$, which completes the proof.

\selectlanguage{english}%
\bibliography{General_Bias_NESS_Entanglement}

\end{document}